\documentclass[3p,times,11pt,review]{elsarticle}

\usepackage{lineno,hyperref,amsmath}

\usepackage{subfigure}
\usepackage{tabularx}
\usepackage{multirow}
\usepackage{booktabs}
\usepackage{graphicx}
\usepackage{epstopdf}

\biboptions{authoryear}

\journal{Journal of Computational Physics}

\begin{document}

\begin{frontmatter}

\title{A high-fidelity finite volume scheme for ideal magnetohydrodynamics equations using boundary variation diminishing algorithm}

\author[XJTU]{Chenxi Pan}

\author[XJTU]{Sheng Song}

\author[XJTU]{Chungang Chen\corref{correspondingauthor}}
\cortext[correspondingauthor]{Corresponding author}
\ead{cgchen@xjtu.edu.cn}

\author[CEMC]{Xingliang Li}

\author[CEMC]{Xueshun Shen}

\author[Titech]{Feng Xiao}

\address[XJTU]{State Key Laboratory for Strength and Vibration of Mechanical Structures and Department of Mechanics, Xi'an Jiaotong University, Xi'an, China}

\address[CEMC]{Center for Earth System Modelling and Predication, China Meteorological Administration, Beijing, China}

\address[Titech]{Department of Mechanical Engineering, Tokyo Institute of Technology, Tokyo, Japan}

\begin{abstract}
A high-fidelity finite volume scheme based on the BVD (boundary variation diminishing) concept is proposed in this study to solve the ideal magnetohydrodynamics (MHD) equations. A hybrid spatial reconstruction profile, consisting of a quadratic polynomial and a steepness-adjustable hyperbolic tangent function, is adopted to reproduce the accurate solutions of the complex magnetohydrodynamics flows. The BVD principle is used to find a optimal combination of these two types of spatial reconstructions by comparing the variations of the interface values interpolated in two adjacent cells, aiming to remove the non-physical oscillations around discontinuities by switching the quadratic polynomial to a step-shaped function. Additionally, a constrained transport (CT) method is applied in this study to assure the non-divergent solution of the magnetic field. The widely used numerical tests in one and two dimensional cases were checked in this study. The numerical results can retrieve the accuracy of a $3^{rd}$-order linear scheme in the convergence test for both advection and MHD equations and capture the strong shock waves in MHD flows without spurious oscillations. In comparison with the results of a $3^{rd}$-order WENO (weighted essentially non-oscillatory) scheme, the proposed one gains more accurate solutions not only for the strong discontinuities but also the smooth structures across scales. To our knowledge, this is the first attempt to build a high-fidelity model for ideal MHD equations by a BVD algorithm. Numerical results are competitive to those of existing advanced schemes and thus the BVD algorithm has promising potentials to build practical models for various MHD flows.

\end{abstract}


\end{frontmatter}

\section{Introduction}\label{sec1}

Many physical phenomena in geophysics, astrophysics, fission and fusion, metallurgy, direct energy conversion and so on are dominated by the magnetohydrodynamics (MHD), which studies the dynamics of an electrically-conducting fluid in a magnetic field. To provide the numerical framework for building simulation models in above fields, it has gained extensive attention to develop the accurate numerical schemes for the ideal MHD equations. However, many difficulties have been encountered since these equations are highly nonlinear and the solutions are characterized by the wave propagations covering a wide range of the spatial and temporal scales. So far, it is still a challenging work in CFD community to reproduce the high-fidelity numerical solutions of the MHD equations.

Many efforts have been made to construct the MHD solver using the high-resolution shock-capturing schemes. \citet{Powell_1999} proposed an solution-adaptive upwind scheme to solve ideal and resistive MHD equations, which is the basic tool to develop the BATS-R-US model in SWMF (space weather modeling framework) software \citep{SWMFReview}. \citet{Pen2003} designed a fast, simple and efficient MHD code using the TVD (total variation diminishing) scheme \citep{LeVeque_2002}. \citet{Stone_2008} developed an open-source code called Athena, which was used for the astrophysical magnetohydrodynamics with the application of the TVD and PPM (piecewise parabolic method) \citep{Colella_1984} schemes.

The high-order essentially non-oscillatory (ENO) and weighted essentially non-oscillatory (WENO) schemes, which has gained great success in building the non-oscillatory numerical models for Euler equations \citep{Harten_1987,Liu_1994,Jiang_1996}, were also used to solve MHD equations. \citet{Jiang_1999} extended the ENO/WENO schemes to MHD equations. \citet{Li_2008} used a non-oscillatory central finite-volume scheme \citep{Liu_2007} to simulate MHD flows. \citet{Shen_2012} developed a MHD model with the application of the high-order WENO scheme and an E-CUSP (energy-convective upwind and split pressure) method. \citet{Ivan_2013} designed a parallel code for Euler and MHD equations on AMR (adaptive mesh refinement) cube-sphere grid using central ENO finite-volume method. More recently, \citet{Liu_2021} proposed a conservative WLS-ENO (weighted-least-squares based essentially non-oscillatory) reconstruction profile and implemented a new MHD model; \cite{Fu_2022} checked the performance of recently developed high-order TENO (targeted essentially non-oscillatory) schemes in a MHD solver.

In order to improve the solution fidelity of non-oscillatory schemes, a new hybrid spatial reconstruction strategy was proposed based on a boundary variation diminishing (BVD) algorithm \citep{Sun_2016,Deng_2018}. To implement a BVD scheme, two or more types of functions are considered as the candidates for the spatial reconstruction, including the high-order polynomials, the step-shaped function, the existing non-oscillatory reconstruction profiles like WENO, TVD and so on. The optimal combination of functions are then determined by minimizing the jump between two estimations of interface values from different types of reconstruction functions. According to the numerical results of Euler equations, the BVD scheme can retrieve the precision of high-order linear schemes in convergence tests and has better fidelity in reproducing the structures across a wide range of spatial scales in comparison with the WENO scheme of the same order. In \citet{Deng_2019}, a $5^{th}$-order BVD scheme, named ${\rm P}_4{\rm T}_2$ scheme, was proposed by using a $4^{th}$-order polynomial and a step-shaped steepness-adjustable hyperbolic tangent function which was originally devised for the interface capturing of multiphase flows. In the benchmark tests of advection and Euler equations, numerical results of ${\rm P}_4{\rm T}_2$ scheme has shown its excellent skills to produce high-fidelity solutions in comparison with existing advanced high-resolution schemes.

In this study, we are going to implement a high-fidelity numerical scheme for ideal MHD equations using the BVD algorithm. The hybrid spatial reconstruction will be accomplished following the numerical framework proposed in \cite{Deng_2019}. Though the $5^{th}$- or higher-order BVD schemes were extensively studied and reported in \cite{Deng_2019,Deng_2020,Tann_2020,Chamarthi_2021} among others, we try to construct $3^{rd}$-order scheme by combining the steepness-adjustable hyperbolic tangent function with a quadratic polynomial based on the BVD concept in this study. To accomplish the spatial reconstruction with a compact stencil, we tend to propose not only accurate but also easy-to-implement numerical scheme for the practical applications using complex computational meshes, e.g., an efficient MHD model with the application of the AMR (adaptive mesh refinement) technique.

The rest part of this paper is organized as follows. Section \ref{sec:Numerical methods} describes the numerical formulations of the proposed MHD model using a two-step BVD scheme. In section \ref{sec:Tests and results}, the benchmark tests for both advection and MHD equations are checked to verify the solution fidelity of the proposed model. The numerical results will be evaluated by compared with those of a $3^{rd}$-order WENO scheme. Finally, a short summary and outlook of the future work will be given in section \ref{sec:conclusions}.

\section{Numerical formulations}\label{sec:Numerical methods}

\subsection{Ideal MHD equations}

In this study, a finite volume scheme is proposed to solve the ideal MHD equations
\begin{eqnarray}\label{MHD}
\frac{\partial{\rho}}{\partial{t}} + \nabla \cdot\left(\rho \boldsymbol{v}\right) &=&0 \\
\frac{\partial{\left(\rho \boldsymbol{v}\right)}}{\partial{t}} + \nabla \cdot \left(\rho \boldsymbol{v} \otimes \boldsymbol{v} +p_t \boldsymbol{I} -\boldsymbol{B} \otimes \boldsymbol{B} \right) &=&0\nonumber \\
\frac{\partial{ \boldsymbol{B}}}{\partial{t}} + \nabla \cdot \left( \boldsymbol{v}\otimes\boldsymbol{B} - \boldsymbol{B}\otimes\boldsymbol{v}\right)&=&0\nonumber\\
\frac{\partial{E}}{\partial{t}} + \nabla \cdot\left[ \left(E+p_t\right)\boldsymbol{v} -\left( \boldsymbol{v} \cdot \boldsymbol{B} \right)\boldsymbol{B} \right] &=&0 \nonumber
\end{eqnarray}
with the non-divergent constraint for the magnetic field
\begin{equation}\label{divergence-free}
\boldsymbol{\nabla} \cdot \boldsymbol{B}=0,
\end{equation}
where $\rho$, $\boldsymbol{v}=\left(u,v,w\right)$ and $\boldsymbol{B}=\left(B_x,B_y,B_z\right)$ denote the density, the velocity vector and the magnetic field (magnetic flux density), the total energy density $E$ is defined as
\begin{equation}\label{EnergyDensity}
E=\frac{p}{\gamma-1} + \rho \frac{1}{2}\boldsymbol{v} \cdot \boldsymbol{v} +\frac{1}{2}\boldsymbol{B} \cdot \boldsymbol{B},
\end{equation}
$p$ and $\gamma$ are the thermodynamic pressure and the ratio of specific heats, $p_t =p+\frac{1}{2}\boldsymbol{B} \cdot \boldsymbol{B}$.

On the Cartesian grid, above MHD equations can be recast into a flux-form for the development of a conservative model as
\begin{equation} \label{MHDConservationLaw}
\frac{\partial\boldsymbol{U}}{\partial t} + \frac{\partial\boldsymbol{F}}{\partial x} + \frac{\partial\boldsymbol{G}}{\partial y} + \frac{\partial\boldsymbol{H}}{\partial z} =0,
\end{equation}
where the dependent variables are $\boldsymbol{U}=\left[\rho,\rho u,\rho v,\rho w, B_x,B_y,B_z,E\right]^T$ and the flux functions in different directions are written as
\begin{equation}
\boldsymbol{F} =\left[
\begin{array}{c}
\rho u \\ \rho u^2+p_t-B^{2}_{x} \\ \rho uv-B_xB_y \\ \rho uw-B_xB_z \\0 \\uB_y-vB_x \\uB_z-wB_x \\ \left(E+p_t\right)u-\left(\boldsymbol{v} \cdot  \boldsymbol{B} \right)B_x
\end{array}
\right],
\boldsymbol{G}  =\left[
\begin{array}{c}
\rho v \\\rho uv-B_y B_x \\ \rho v^2+p_t-B^{2}_{y} \\\rho vw-B_yB_z \\vB_x-uB_y \\0 \\vB_z-wB_y \\ \left(E+p_t\right)v-\left(\boldsymbol{v} \cdot \boldsymbol{B}\right)B_y
\end{array}
\right]\ \rm{and}\
\boldsymbol{H} =\left[
\begin{array}{c}
\rho w \\ \rho uw -B_zB_x \\ \rho vw-B_zB_y \\ \rho w^2+p_t-B^{2}_z \\wB_x-uB_z \\wB_y-vB_z \\0 \\ \left(E+p_t\right)w-\left(\boldsymbol{v} \cdot \boldsymbol{B}\right)B_z
\end{array}
\right].
\end{equation}

\subsection{BVD scheme for one-dimensional scalar equation}

The BVD algorithm is adopted in this study to build a hybrid spatial reconstruction profile for the implementation of a high-fidelity solver of the ideal MHD equations. We first describe the numerical formulations of a two-step BVD algorithm considering the one-dimensional scalar conservation law written in the flux-form as
\begin{equation}\label{ScalarConservationLaw}
\frac{ \partial q }{ \partial t } + \frac{ \partial f\left(q\right) }{ \partial x }=0,
\end{equation}
where $q(x,t)$ is the dependent variable and $f \left(q\right)$ is the flux function.

We divide the computational domain $x\in\left[x_l,x_r\right]$ into $N$ non-overlapping uniform cells and grid spacing is $\Delta x=\frac{x_r-x_l}{N}$. Using the finite volume method, a semi-discrete form of Eq. \eqref{ScalarConservationLaw} is obtained as
\begin{equation}\label{SemiDiscreteODE}
\frac{{\rm d} q_i(t)}{{\rm d}t}=-\frac{1}{\Delta x}\left(\hat{f}_{i+\frac{1}{2}}-\hat{f}_{i-\frac{1}{2}}\right),
\end{equation}
where $q_i(t)$ is the volume-integrated average of variable $q\left(x,t\right) $ over the cell $\mathcal{C}_i$ $\left(x\in\left[x_{i-\frac{1}{2}},x_{i+\frac{1}{2}}\right]\right)$ at time $t$, i.e.
\begin{equation}\label{IntAve}
q_i(t) = \frac{1}{\Delta x} \int_{x_{i-\frac{1}{2}}}^{x_{i+\frac{1}{2}}} q\left(x,t\right)\mathrm{d}x,
\end{equation}
and $\hat{f}_{i\pm\frac{1}{2}}$ is approximations of the fluxes across the cell interfaces $x=x_{i\pm\frac{1}{2}}$.

To build an upwind scheme, the numerical flux $\hat{f}_{i+\frac{1}{2}}$ is obtained by solving a Riemann problem as
\begin{equation}\label{RiemannSolver}
\hat{f}_{i+\frac{1}{2}} = {\rm Riemann} \left( f\left(q^L_{i+\frac{1}{2}}\right) , f\left(q^R_{i+\frac{1}{2}}\right) \right),
\end{equation}
where $q^L_{i+\frac{1}{2}}$, $q^R_{i+\frac{1}{2}}$ are values of dependent variable at cell interface $\left(x=x_{i+\frac{1}{2}}\right)$ obtained through the spatial reconstructions of two adjacent cells $\mathcal{C}_i$ and $\mathcal{C}_{i+1}$.

The ordinary differential equation (ODE) \eqref{SemiDiscreteODE} is then solved by an explicit third-order TVD Runge-Kutta method \citep{Shu_1988} in this study. Provided the known solution at $t=t^n$, the numerical solution at the next time step is approximated by
\begin{eqnarray}\label{RK3}
q^{\left(1\right)} &=& q^n +\Delta t \mathcal{D}\left(q^n\right), \\
q^{\left(2\right)} &=& \frac{3}{4} q^n +\frac{1}{4} q^{\left(1\right)} +\frac{1}{4} \Delta t \mathcal{D}\left(q^{(1)}\right), \nonumber \\
q^{n+1} &=& \frac{1}{3} q^n +\frac{2}{3} q^{\left(2\right)} +\frac{2}{3} \Delta t \mathcal{D}\left(q^{\left(2\right)}\right),\nonumber
\end{eqnarray}
where the operator $\mathcal{D}$ represents the finite-volume spatial discretization. In this study, the CFL number of 0.4 number is used for all tests .

The key task left here is to accomplish the piecewise spatial reconstructions for estimating values of the dependent variable $q$ and the flux function $f$ at cell interfaces. Using the BVD concept, several alternative functions having different forms are carefully chosen to minimize the boundary variations in order to build an accurate and non-oscillatory scheme. The numerical formulations of a $3^{th}$-order finite volume scheme using BVD algorithm are described as follows.

\subsubsection{Functions for spatial reconstruction}

Here, we introduce two alternative functions of different types for the spatial reconstruction in this study. It should be pointed out that the BVD concept provides a very flexible framework to devise the high-fidelity finite volume models for flows across a wide range of spatial scales, it is still an open question and worth further investigations on the choice of the alternative reconstruction functions, in order to develop the accurate numerical schemes in various applications.

\begin{itemize}

\item \emph{A quadratic polynomial}

A $3^{th}$-order finite volume scheme can be constructed with the application of a piecewise quadratic interpolation polynomial, i.e.
\begin{equation}\label{2ndpoly}
\displaystyle Q^P_i\left(x\right)=\sum_{s=0}^2c_i^s\left(x-x_i\right)^s
\end{equation}
for the cell $\mathcal{C}_i$, where the coefficients $c^s_i\ \left(s=0\ {\rm to}\ 2\right)$ are determined through following constraints
\begin{equation}\label{PolyInt}
\int_{x_{k-\frac{1}{2}}}^{x_{k+\frac{1}{2}}}Q^P_i\left(x\right)\mathrm{d}x=q_k\ \left(k=i-1,i\ {\rm and}\ i+1\right).
\end{equation}

Using the reconstruction polynomial Eq. \eqref{2ndpoly}, the values of dependent variable at cell interfaces can be approximated as
\begin{eqnarray}
\left(\hat{q}^R_{i-\frac{1}{2}}\right)^P&=&  \frac{1}{3}q_{i-1} +\frac{5}{6}q_{i} -\frac{1}{6}q_{i+1},\\
\left(\hat{q}^L_{i+\frac{1}{2}}\right)^P&=& -\frac{1}{6}q_{i-1} +\frac{5}{6}q_{i} + \frac{1}{3}q_{i+1}.
\end{eqnarray}

\item \emph{A steepness-adjustable hyperbolic tangent function}

A hyperbolic tangent function was devised in \cite{Xiao_2005,Xiao_2011} to develop an accurate VOF-type interface capturing method, called THINC (tangent hyperbola for interface capturing). It is step-shaped and expressed within the cell $\mathcal{C}_i$ as
\begin{equation}\label{THINC}
Q_i^{T}\left(x\right) =m+\frac{M}{2}\left\{1+\Theta \  \mathrm{tanh}\left[\beta \left(\frac{x-x_{i-\frac{1}{2}}}{\Delta x}-c\right)\right]  \right\}.
\end{equation}

The parameters in Eq. \eqref{THINC} are defined as $m=\min\left(q_{i-1},q_{i+1}\right)$, $M=\max\left(q_{i-1},q_{i+1}\right)$, $\Theta=\mathrm{sign}\left(q_{i+1}-q_{i-1}\right)$ and parameter $c$ is adopted to preserve the cell-integrated average of the variable $q$ over the cell $\mathcal{C}_i$. Different values can be assigned to parameter $\beta$ in order to adjust the steepness of the jump within the cell $\mathcal{C}_i$.

The values at cell interfaces interpolated using above hyperbolic tangent function are
\begin{eqnarray}
\left(\hat{q}^{R}_{i-\frac{1}{2}}\right)^T &=&m+\frac{M}{2}\left(1+\Theta  A  \right), \\
\left(\hat{q}^{L}_{i+\frac{1}{2}}\right)^T &=&m+\frac{M}{2}\left(1+\Theta  \frac{\tanh\beta+A}{1+A  \tanh\beta}  \right),
\end{eqnarray}
where $A=\frac{B-\cosh\beta}{\sinh\beta}$, $B=\exp \left[\Theta\beta \left( 2\frac{q_{i}-m+\epsilon}{M+\epsilon}-1 \right)\right]$ and $\epsilon=10^{-20}$.

Additionally, a piecewise constant function is used to remove the non-physical oscillations if $ (q_{i+1}-q_{i})(q_{i}-q_{i-1}) \leq 0$, i.e. $\left(\hat{q}^{R}_{i-\frac{1}{2}}\right)^T=\left(\hat{q}^{L}_{i+\frac{1}{2}}\right)^T=q_i$.

\end{itemize}

Above two interpolation functions are used in this study to build a hybrid spatial reconstruction, which can retrieve the accuracy of $3^{rd}$-order upwind finite volume scheme in smooth regions by using the quadratic polynomial and remove the non-physical oscillations around the discontinuities by switching to the hyperbolic tangent function for the piecewise spatial reconstructions of the ``troubled" cells. The BVD principle is adopted to precisely distinguish the smooth and nonsmooth cells in the numerical simulations.

\subsubsection{BVD algorithm}\label{subsec:BVDalgorithm}

The basic idea of the BVD principle is to choose the proper reconstruction function by minimizing the boundary variations of two unequal interpolated values at cell interface. In this study, we use a two-step BVD framework proposed in \cite{Deng_2019} to develop a $3^{rd}$-order finite volume scheme for solving ideal MHD equations.

A two-step BVD algorithm is conducted with the application of three different types of reconstruction functions, including $Q^P\left(x\right)$, $Q^{T_{1.05}}\left(x\right)$ and $Q^{T_{1.2}}\left(x\right)$, where the superscripts $P$ and $T$ denote the quadratic polynomial and the hyperbolic tangent function. The subscripts $1.05$ and $1.2$ denote the two different values of parameter $\beta$ to change the steepness of the hyperbolic tangent function. In this study, we use different values of $\beta$ in comparison with the ${\rm P}_4{\rm T}_2$ scheme \citep{Deng_2019}, which are tuned for MHD model based on numerical results of benchmark tests given in section 3.

To accomplish a two-step BVD algorithm, we first define the total boundary variation (TBV) for the cell $\mathcal{C}_i$ as
\begin{equation}
{\rm TBV}_i=\left|Q_i\left(x_{i-\frac{1}{2}}\right)-Q_{i-1}\left(x_{i-\frac{1}{2}}\right)\right|+\left|Q_{i+1}\left(x_{i+\frac{1}{2}}\right)-Q_{i}\left(x_{i+\frac{1}{2}}\right)\right|,
\end{equation}
where $Q_i$ and $Q_{i\pm1}$ are the spatial reconstructions for cells $\mathcal{C}_i$ and $\mathcal{C}_{i\pm1}$, which can be one of three candidates $Q^P\left(x\right)$, $Q^{T_{1.05}}\left(x\right)$ and $Q^{T_{1.2}}\left(x\right)$.

By minimizing the TBV of each cell, the optimal interpolation function is determined as follows.

\begin{itemize}

\item[1)]{Step one}

We first consider two functions $Q^P\left(x\right)$ and $Q^{T_{1.05}}\left(x\right)$ to construct a non-oscillatory scheme by checking the total boundary variations ${\rm TBV}^{P}_i$ and ${\rm TBV}^{T_{1.05}}_i$ being calculated using these two reconstructions.

The reconstruction functions for all cells are initialized to use quadratic polynomials defined in \eqref{2ndpoly}, i.e. $Q^{\rm one}_i\left(x\right)=Q^P_i\left(x\right)$. In this step, we revise the spatial reconstruction in cell $\mathcal{C}_{i}$ as
\begin{equation}
\label{BVD1}
Q_i^{\rm one}\left(x\right)= Q_i^{T_{1.05}}\left(x\right),\ {\rm if} \ \exists  \ k \in \left[i-1,i,i+1\right],\ {\rm TBV}_k^{T_{1.05}}<{\rm TBV}_k^{P}.
\end{equation}

\item[2)]{Step two}

The total boundary variations are further optimized by considering functions $Q^{\rm one}$ and $Q^{T_{1.2}}\left(x\right)$, a hyperbolic tangent function with larger value of $\beta$ to reduce the numerical dissipation. The spatial reconstruction in the cell $\mathcal{C}_i$ is finally determined as
\begin{equation}
Q_i^{\rm BVD}\left(x\right)=\left\{
\begin{array}{ll}
Q_i^{T_{1.2}}\left(x\right),  &   {\rm if} \ {\rm TBV}_i^{T_{1.2}} < {\rm TBV}_i^{\rm one}       \\
Q_i^{\rm one}\left(x\right),  &   {\rm otherwise}
\end{array}\right..
\end{equation}

Please note we revise the reconstruction functions only considering the TBVs in cell $\mathcal{C}_i$ in step two. It is helpful to sharpen the solutions of discontinuities without generating extra numerical oscillations.

\end{itemize}

\subsection{BVD scheme for MHD equations}\label{subsec:ExtensionToMHD}

In this section, we describe the numerical procedure to solve the MHD equations expressed in a flux form in Eq. \eqref{MHDConservationLaw}.

Using a finite volume method, a semi-discrete form of Eq. \eqref{MHDConservationLaw} is obtained as
\begin{equation}\label{SemiDiscreteMHD}
\frac{{\rm d} \boldsymbol{U}_{ij}}{{\rm d}t} =
-\frac{1}{\Delta x} \left(\hat{\boldsymbol{F}}_{i+\frac{1}{2}j} - \hat{\boldsymbol{F}}_{i-\frac{1}{2}j}\right)
-\frac{1}{\Delta y} \left(\hat{\boldsymbol{G}}_{ij+\frac{1}{2}} - \hat{\boldsymbol{G}}_{ij-\frac{1}{2}}\right).
\end{equation}

In this study, the two-dimensional MHD problems are solved to verify the $3^{rd}$-order BVD scheme, i.e. model variables are uniformly distributed in $z$-direction. Thus, the derivatives of flux functions $\boldsymbol{H}\left(\boldsymbol{U}\right)$ are always zero and therefore neglected in Eq. \eqref{SemiDiscreteMHD}. Numerical fluxes $\hat{\boldsymbol{G}}$ and $\hat{\boldsymbol{H}}$ across the cell boundaries are estimated by applying above one dimensional interpolation operation in different directions one by one. Strictly speaking, the resulting scheme has $2^{nd}$-order accuracy at most. But this strategy is more efficient and widely adopted by the practical finite volume models.

In this study, a simple local Lax-Friedrichs Riemann solver is used build an efficient upwind scheme. For example, the Riemann problem in $x$-direction at $x=x_{i+\frac{1}{2}}$ is solved by
\begin{equation}\label{LLF}
\hat{\boldsymbol{F}}_{i+\frac{1}{2}j} =\frac{1}{2} \left[
\boldsymbol{F}\left(\hat{\boldsymbol{U}}^L_{i+\frac{1}{2}j}\right) +
\boldsymbol{F}\left(\hat{\boldsymbol{U}}^R_{i+\frac{1}{2}j}\right)
+\alpha_{i+\frac{1}{2}j}\left(\hat{\boldsymbol{U}}^L_{i+\frac{1}{2}j} - \hat{\boldsymbol{U}}^R_{i+\frac{1}{2}j} \right)
\right],
\end{equation}
where $\hat{\boldsymbol{U}}^L_{i+\frac{1}{2}}$ and $\hat{\boldsymbol{U}}^R_{i+\frac{1}{2}}$ are estimations of dependent variables at cell boundary $x=x_{i+\frac{1}{2}}$ by $x$-direction spatial reconstruction functions of cells $\mathcal{C}_{ij}$ and $\mathcal{C}_{i+1j}$ respectively, parameter $\alpha_{i+\frac{1}{2}j}=\left|u_{i+\frac{1}{2}j}\right|+{c_f}_{i+\frac{1}{2}j}$ is the maximum wave speed of a MHD system with $c_f$ being the speed of fast magnetosonic wave
\begin{equation}
c_f=\frac{\sqrt{2}}{2}\sqrt{a^2+b^2+\sqrt{\left(a^2+b^2\right)^2-4a^2C_a^2}},
\end{equation}
and $a=\sqrt{\frac{\gamma p}{\rho}}$ is the sound speed, $C_a=\sqrt{\frac{B_x^2}{\rho}}$ is the speed of Alfv\'{e}n wave, $b=\sqrt{\frac{B_x^2+B_y^2+B_z^2}{\rho}}$.

In $y$-direction, the numerical fluxes $\hat{\boldsymbol{G}}_{ij\pm\frac{1}{2}}$ can be determined in the similar way.

\subsubsection{BVD spatial reconstruction in MHD solver}

To remove the non-physical oscillations around discontinuities, the BVD algorithm is used to build the piecewise spatial construction. To achieve accurate results for a hyperbolic system, the BVD algorithm is applied considering the characteristic variables in a MHD solver.

To evaluate numerical flux $\hat{\boldsymbol{F}}$ by the known solution of dependent (conservative) variables, the following numerical operations are applied to interpolate $\hat{\boldsymbol{U}}^R_{i-\frac{1}{2}j}$ and $\hat{\boldsymbol{U}}^L_{i+\frac{1}{2}j}$ by the spatial reconstruction function of the cell $\mathcal{C}_{ij}$ in $x$-direction. For the sake of brevity, the subscript $j$ is omitted in the following description.

\begin{itemize}
\item[1)]
Compute the primitive variables $\boldsymbol{u}=\left[\rho,u,v,w,B_x,B_y,B_z,p\right]^T$ for cells $\mathcal{C}_{i-1}$, $\mathcal{C}_i$ and $\mathcal{C}_{i+1}$ from the known conservative (dependent) variables $\boldsymbol{U}$.
	
\item[2)]
Evaluate the primitive variables at cell boundaries by averaging their values of two adjacent cells as
\begin{equation}\label{MidState}
\boldsymbol{u}_{i-\frac{1}{2}}=\frac{1}{2}\left(\boldsymbol{u}_{i-1}+\boldsymbol{u}_{i}\right)\ \rm{and}\ \boldsymbol{u}_{i+\frac{1}{2}}=\frac{1}{2}\left(\boldsymbol{u}_{i}+\boldsymbol{u}_{i+1}\right).
\end{equation}

\item[3)]
Determine the right and left eigenvectors of Jacobian matrix at $x=x_{i\pm\frac{1}{2}}$ which are used for compute the characteristic variables for BVD reconstruction as
\begin{equation}\label{ComputeEigenvectors}
\boldsymbol{\mathcal{R}}_{i\pm\frac{1}{2}}=\boldsymbol{\mathcal{R}}\left( \boldsymbol{u}_{i\pm\frac{1}{2}} \right)\ \  \rm{and} \ \
\boldsymbol{\mathcal{L}}_{i\pm\frac{1}{2}}=\boldsymbol{\mathcal{L}} \left( \boldsymbol{u}_{i\pm\frac{1}{2}} \right)	
\end{equation}

For the sake of brevity, the detailed expressions of matrices $\boldsymbol{\mathcal{R}}$ and $\boldsymbol{\mathcal{L}}$ are not given in this study and can be referred to \cite{Powell_1999}.

\item[4)]
Calculate values of  characteristic variables by
\begin{equation}\label{PrimToChara}
\boldsymbol{W}_{k}^{L}=\boldsymbol{\mathcal{L}}_{i-\frac{1}{2}} \boldsymbol{u}_{k} \ {\rm and} \
\boldsymbol{W}_{k}^{R}=\boldsymbol{\mathcal{L}}_{i+\frac{1}{2}} \boldsymbol{u}_{k}
\end{equation}
where $k=i-1,\ i\ {\rm and}\ i+1$.

Values of characteristic variables are evaluated in three immediate cells according to the spatial reconstruction stencil used in this study. Please note that there are two sets of characteristic variables (denoted by superscripts $L$ and $R$) used for building two spatial reconstruction functions within cell $\mathcal{C}_i$, which are used to evaluate the interface values at left and right cell boundaries respectively.

\item[5)]
Determine the interface values of characteristic variables using two-step BVD algorithm described above
\begin{eqnarray}
\hat{\boldsymbol{W}}_{i-\frac{1}{2}}^R&=&{\boldsymbol{\mathcal{W}}^L}_i^{\rm BVD}\left(x_{i-\frac{1}{2}}\right),\\
\hat{\boldsymbol{W}}_{i+\frac{1}{2}}^L&=&{\boldsymbol{\mathcal{W}}^R}_i^{\rm BVD}\left(x_{i+\frac{1}{2}}\right),
\end{eqnarray}
where ${\boldsymbol{\mathcal{W}}^L}_i^{\rm BVD}$ and ${\boldsymbol{\mathcal{W}}^R}_i^{\rm BVD}$ are optimal reconstructions determined by the BVD algorithm based on two sets of characteristic variables calculated in the last step.

\item[6)]
Calculate the primitive variables at cell boundaries as
\begin{eqnarray}\label{Prim}
\hat{\boldsymbol{u}}_{i-\frac{1}{2}}^{R}&=&\boldsymbol{\mathcal{R}}_{i-\frac{1}{2}}\hat{\boldsymbol{W}}_{i-\frac{1}{2}}^{R} \\
\hat{\boldsymbol{u}}_{i+\frac{1}{2}}^{L}&=&\boldsymbol{\mathcal{R}}_{i+\frac{1}{2}} \hat{\boldsymbol{W}}_{i+\frac{1}{2}}^{L}
\end{eqnarray}

Finally, the values of conservative variables at cell boundaries, i.e. $\hat{\boldsymbol{U}}_{i-\frac{1}{2}}^{R}$ and $\hat{\boldsymbol{U}}_{i+\frac{1}{2}}^{L}$ can be obtained straightforwardly to solve the Riemann problem defined in Eq. \eqref{LLF}.

\end{itemize}

\subsubsection{Non-divergent correction for magnetic field}

Above numerical formulations can not strictly preserve the divergence-free property of the magnetic field. The divergence error of the magnetic field may lead to the non-physical oscillations and even the numerical instability of a MHD solver \citep{ROSSMANITH_2006}. Thus, it is of essential importance for a MHD model to involve a correction algorithm for rigorously assuring the non-divergent solution of the magnetic field. Several numerical techniques have been proposed to remove the divergence error.
Powell's eight-wave method \citep{Powell_1999} adds the source item proportional to the magnitude of $\boldsymbol{\nabla} \cdot \boldsymbol{B}$ in MHD equations to remove the divergence error of the magnetic field. The constrained transport (CT) method \citep{Yee_1966,Evans_1988,Balsara_1999,Gardiner_2005}
solves the magnetic field written in the curl form instead of the divergence form
to strictly preserve $\boldsymbol{\nabla} \cdot \boldsymbol{B}=0$. The correction technique based on the Poisson projection method \citep{Brackbill_1980,Ryu_1995} was also widely used. However, it has restrictions on the choice of boundary conditions. In this paper, we use the correction algorithm proposed by \cite{Balsara_1999} since it is very efficient and easy-to-implement for the finite volume MHD solvers on structured grids. We briefly describe the numerical procedure for the non-divergent correction as follows.

Considering the third equation of Eq. \eqref{MHD}, so called Faraday's Law, it is recast into a curl form as
\begin{equation}\label{FaradayLaw}
\frac{\partial{ \boldsymbol{B}}}{\partial{t}} + \boldsymbol{\nabla} \times \boldsymbol{\mathcal{E}}  = 0,
\end{equation}
where the electric field $\boldsymbol{\mathcal{E}}$ is
\begin{equation}\label{ElectricField}
\boldsymbol{\mathcal{E}}=-\boldsymbol{v} \times\boldsymbol{B}.
\end{equation}

Taking the divergence of Eq. \eqref{FaradayLaw} and considering  the relation $\boldsymbol{\nabla}  \cdot (\boldsymbol{\nabla} \times \boldsymbol{\mathcal{E}})\equiv 0$, we have
\begin{equation}\label{DivergenceFree}
\frac{\partial{ \left(\boldsymbol{\nabla}  \cdot \boldsymbol{B}\right)} }{\partial{t}}=0,
\end{equation}
which means the magnetic field always satisfies $\boldsymbol{\nabla}  \cdot \boldsymbol{\rm{B}}=0$
if a non-divergent initial condition is specified.

The following relations hold between the flux functions of MHD equations and the components of the electric field,
\begin{equation}\label{ElectricFieldAndFluxs}
\mathcal{E}_x=H_6=-G_7, \ \mathcal{E}_y=F_7=-H_5, {\rm and} \ \mathcal{E}_z=G_5=-F_6.
\end{equation}

A staggered mesh magnetic field transport algorithm \citep{Balsara_1999} is adopted to correct the divergence error generated by above finite volume solver. A auxiliary magnetic field ($\boldsymbol{\mathcal{B}}$) is introduced to accomplish the divergence correction. Considering a cubic cell $\mathcal{C}_{ijk}$, the auxiliary variable $\boldsymbol{\mathcal{B}}$ is defined in a staggered way by using surface-integrated averages of x, y and z components over six boundary surfaces. In $x$-direction, x component $\mathcal{B}_x$ is arranged at two surfaces perpendicular to the x-axis as ${\mathcal{B}_x}_{i\pm\frac{1}{2}jk}$. Similarly, ${\mathcal{B}_y}_{ij\pm\frac{1}{2}k}$ and ${\mathcal{B}_z}_{ijk\pm\frac{1}{2}}$ are defined at boundary surfaces perpendicular to the $y$ and $z$-axes to specify y component $\mathcal{B}_y$ and z component $\mathcal{B}_z$ respectively.

Solving Eq. \eqref{FaradayLaw} by a finite volume scheme, we have the time tendencies of the auxiliary magnetic field as
\begin{eqnarray}
\label{semi-discrete-Bx}\frac{\rm d}{{\rm d}t} {\mathcal{B}_x}_{i+\frac{1}{2}jk}&=&
\frac{{\mathcal{E}_y}_{i+\frac{1}{2}jk+\frac{1}{2}}-{\mathcal{E}_y}_{i+\frac{1}{2}jk-\frac{1}{2}}}{\Delta z}-\frac{{\mathcal{E}_z}_{i+\frac{1}{2}j+\frac{1}{2}k}-{\mathcal{E}_z}_{i+\frac{1}{2}j-\frac{1}{2}k}}{\Delta y}, \\
\label{semi-discrete-By}\frac{\rm d}{{\rm d}t} {\mathcal{B}_y}_{ij+\frac{1}{2}k}&=&
\frac{{\mathcal{E}_z}_{i+\frac{1}{2}j+\frac{1}{2}k}-{\mathcal{E}_z}_{i-\frac{1}{2}j+\frac{1}{2}k}}{\Delta x}-\frac{{\mathcal{E}_x}_{ij+\frac{1}{2}k+\frac{1}{2}}-{\mathcal{E}_x}_{ij+\frac{1}{2}k-\frac{1}{2}}}{\Delta z}, \\
\label{semi-discrete-Bz}\frac{\rm d}{{\rm d}t} {\mathcal{B}_z}_{ijk+\frac{1}{2}}&=&
\frac{{\mathcal{E}_x}_{ij+\frac{1}{2}k+\frac{1}{2}}-{\mathcal{E}_x}_{ij-\frac{1}{2}k+\frac{1}{2}}}{\Delta y}-\frac{{\mathcal{E}_y}_{i+\frac{1}{2}jk+\frac{1}{2}}-{\mathcal{E}_y}_{i-\frac{1}{2}jk+\frac{1}{2}}}{\Delta x},
\end{eqnarray}
where line-integrated averages of the electric field along the cell edges give the electromotive force (EMF) and can be estimated through averaging the fluxes across the four boundary surfaces sharing the edge as
\begin{eqnarray}\label{semi-discrete-E}
{\mathcal{E}_x}_{ij+\frac{1}{2}k+\frac{1}{2}}&=&\frac{1}{4}
\left( {H_6}_{ijk+\frac{1}{2}} + {H_6}_{ij+1k+\frac{1}{2}} - {G_7}_{ij+\frac{1}{2}k} - {G_7}_{ij+\frac{1}{2}k+1} \right), \\
{\mathcal{E}_y}_{i+\frac{1}{2}jk+\frac{1}{2}}&=&\frac{1}{4}
\left( {F_7}_{i+\frac{1}{2}jk} + {F_7}_{i+\frac{1}{2}jk+1} - {H_5}_{ijk+\frac{1}{2}} - {H_5}_{i+1jk+\frac{1}{2}} \right), \\
{\mathcal{E}_z}_{i+\frac{1}{2}j+\frac{1}{2}k}&=&\frac{1}{4}
\left( {G_5}_{ij+\frac{1}{2}k} + {G_5}_{i+1j+\frac{1}{2}k} - {F_6}_{i+\frac{1}{2}jk} - {F_6}_{i+\frac{1}{2}j+1k} \right).
\end{eqnarray}

It is easy to prove that the divergence-free property is always preserved for the auxiliary magnetic field $\boldsymbol{\mathcal{B}}$ if the divergence is calculated by
\begin{equation}
\boldsymbol{\nabla}\cdot\boldsymbol{\mathcal{B}}=\frac{{\mathcal{B}_x}_{i+\frac{1}{2}jk}-{\mathcal{B}_x}_{i-\frac{1}{2}jk}}{\Delta x}+\frac{{\mathcal{B}_y}_{ij+\frac{1}{2}jk}-{\mathcal{B}_y}_{ij\-\frac{1}{2}jk}}{\Delta y}+\frac{{\mathcal{B}_z}_{ijk+\frac{1}{2}}-{\mathcal{B}_z}_{ijk-\frac{1}{2}}}{\Delta z}.
\end{equation}

Above correction is applied at the end of each substep of the Runge-Kutta time marching scheme. To obtain a non-divergent solution for the magnetic field, we replace the solution of $\boldsymbol{B}$ by that of the auxiliary variable $\boldsymbol{\mathcal{B}}$ as
\begin{equation}\label{B-corrected}
\boldsymbol{B}_{ijk}\left(t\right)=\boldsymbol{\mathcal{B}}_{ijk}\left(t\right),
\end{equation}
where the volume-integrated average of the auxiliary magnetic field are obtained by averaging the surface-integrated values updated through Eqs. \eqref{semi-discrete-Bx} to \eqref{semi-discrete-Bz}
\begin{eqnarray}
{\mathcal{B}_x}_{ijk}&=&\frac{1}{2}\left({\mathcal{B}_x}_{i-\frac{1}{2}jk}+{\mathcal{B}_x}_{i+\frac{1}{2}jk}\right), \\
{\mathcal{B}_y}_{ijk}&=&\frac{1}{2}\left({\mathcal{B}_y}_{ij-\frac{1}{2}k}+{\mathcal{B}_y}_{ij+\frac{1}{2}k}\right), \\
{\mathcal{B}_z}_{ijk}&=&\frac{1}{2}\left({\mathcal{B}_z}_{ijk-\frac{1}{2}}+{\mathcal{B}_z}_{ijk+\frac{1}{2}}\right).
\end{eqnarray}

Additionally, an optional step can be adopted to revise the total energy density by
\begin{equation}\label{Energy-corrected}
E=E+\frac{1}{2}\left( \boldsymbol{\mathcal{B}}\cdot\boldsymbol{\mathcal{B}} - \boldsymbol{B}\cdot\boldsymbol{B} \right).
\end{equation}

This correction helps to assure the positivity of the pressure in numerical simulations, but at a cost of losing the rigorous energy conservation.

\section{Numerical tests and results}\label{sec:Tests and results}

In this section, the proposed numerical scheme is checked by the linear advection equation and MHD equations. The tests with smooth solutions are used to verify the accuracy of the proposed numerical scheme, which is expected to achieve not only the convergence rate but also the magnitude of errors in comparison with the $3^{rd}$-order linear scheme. Other tests including both complex multi-scale structures and discontinuities are used to assess the solution fidelity of the proposed BVD scheme.

\subsection{Test cases for linear advection equation}

\subsubsection{Advection of a sine wave }

The convergence rate of the proposed scheme is first checked on a series of refining grids. The initial condition is given as
\begin{equation}\label{InitSin}
q(x,0)=\sin\left(\pi x\right) , x \in \left[ -1 , 1 \right],
\end{equation}
with a constant advection speed of $u=1$.

The computational errors $l_1$ and $l_\infty$ at t = 2 (after one complete period) are examined, which are defined by
\begin{equation}\label{error norms}
\displaystyle l_1=\frac{1}{N}\sum_{i=1}^N \left| {q_e}_i - q_i \right|
\ {\rm and}\
l_{\infty}= \max_{1 \leqslant i \leqslant N} \left| {q_e}_i - q_i \right|,
\end{equation}
where ${q_e}$ is the exact solution and $q$ the numerical one.

The proposed BVD scheme is tested using two settings, i.e. the spatial reconstructions are accomplished by just step one or complete two-step BVD algorithm as described in section 2.2.2. The numerical results of a $3^{rd}$-order WENO scheme (denoted hereafter by WENO3) and a linear upwind one (denoted hereafter by upwind3) are also checked for comparison. The computational errors and corresponding convergence rates are given in Table \ref{accuracy}.

One-step and two-step BVD schemes achieve the identical accuracy as the linear upwind scheme in this considerably smooth problem. It reveals that the BVD algorithm always choose the quadratic polynomial as the optimal function for the spatial reconstructions in this tests. The $3^{rd}$-order WENO scheme shows relatively low accuracy in this tests as it was found in \cite{Acker_2016} that WENO schemes may treat some smooth regions as discontinuities in the case of relatively coarse grids. Though WENO3 has faster convergence rates on two fine grids, the $l_1$ and $l_\infty$ errors are one and two orders of magnitude larger than other three schemes on the finest grid.

\subsubsection{Advection of a smooth profile containing critical points}\label{subsubsec:AccuracyTestOfASmoothProfileContainingCriticalPoints}

A test case containing some critical points, where high-order derivatives do not simultaneously vanish, was proposed in \cite{Henrick_2005}. It is more challenging for numerical schemes to distinguish smooth and non-smooth cells in the numerical simulations. The initial condition is given by
\begin{equation}\label{InitSin4}
q(x,0)=\sin^4\left(\pi x\right) , x \in \left[ -1 , 1 \right].
\end{equation}

\begin{figure}[htbp]
\centering
\subfigure[$l_1$ error]
{\includegraphics[width=0.45\textwidth]{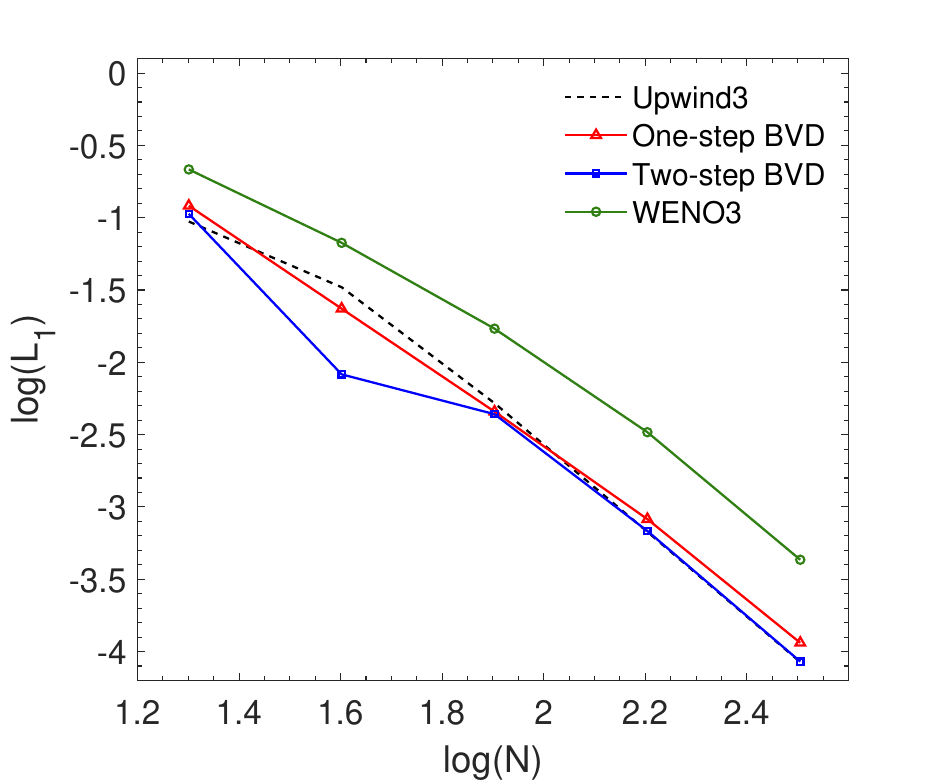}}
\subfigure[$l_\infty$ error]
{\includegraphics[width=0.45\textwidth]{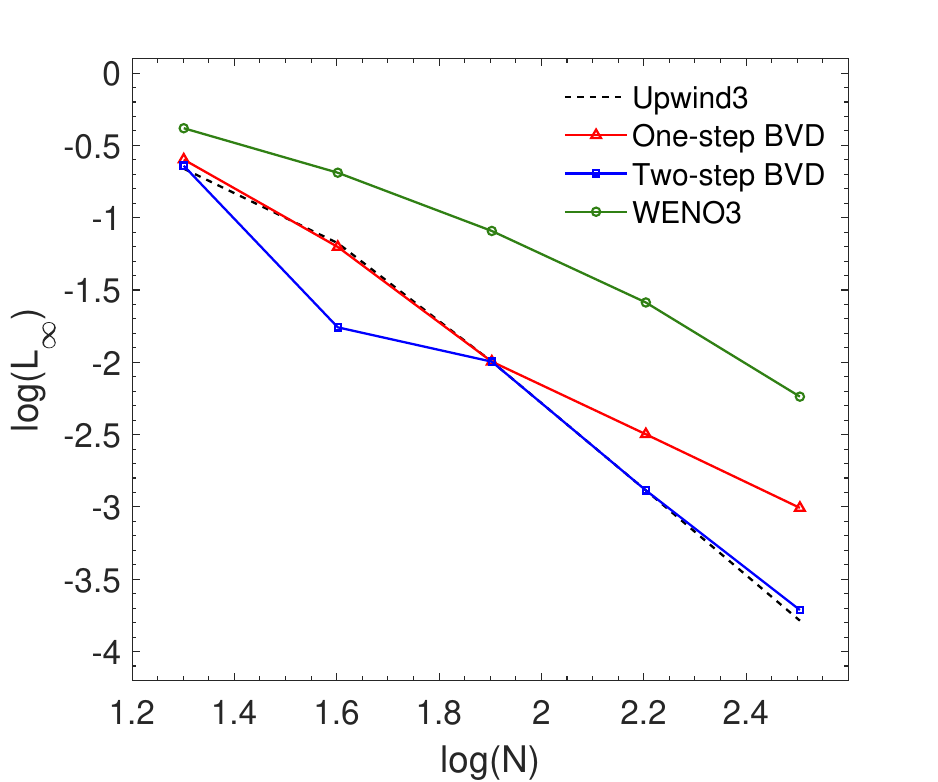}}
\caption{Errors of advection of a smooth profile containing critical points on a series of refining  grids at $t=2$ by different schemes.}
\label{ErrorsSin4}
\end{figure}

We plot $l_{1}$ and $l_{\infty}$ errors at $t=2$ on a series of refining grids for WENO3 and the proposed BVD schemes in comparison with linear upwind3 scheme in Fig. \ref{ErrorsSin4}. In this test, four schemes give the different results. It means both BVD and WENO scheme treat the solutions as discontinuities in some cells. Considering the $l_1$ error (left panel), the results of BVD schemes with one and two-step reconstructions agree well with that of the linear scheme. Two-step algorithm can improve the numerical results and give visually identical $l_1$ error on two fine grids. The WENO3 scheme achieves the similar convergence rate, but generates obviously larger errors. Only two-step BVD scheme performs well in comparison with upwind3 scheme considering $l_\infty$ errors (right panel). On fine grids, both one-step BVD and WENO3 schemes converge slowly and generate larger errors. But one-step BVD scheme is still more accurate than WENO3 scheme.

\subsubsection{Advection of Jiang and Shu's complex wave }

As a widely used benchmark test, the initial distribution of Jiang and Shu's complex wave consists of four types of profiles involving different characteristics and details can be referred to \cite{Jiang_1996}. Since the solution includes both discontinuous and continuous regions, it has been extensively tested in existing literatures to assess the abilities of the numerical schemes to produce the high-fidelity results.

\begin{figure}[htbp]
\centering
\subfigure[WENO3]	
{\includegraphics[width=0.32\textwidth]{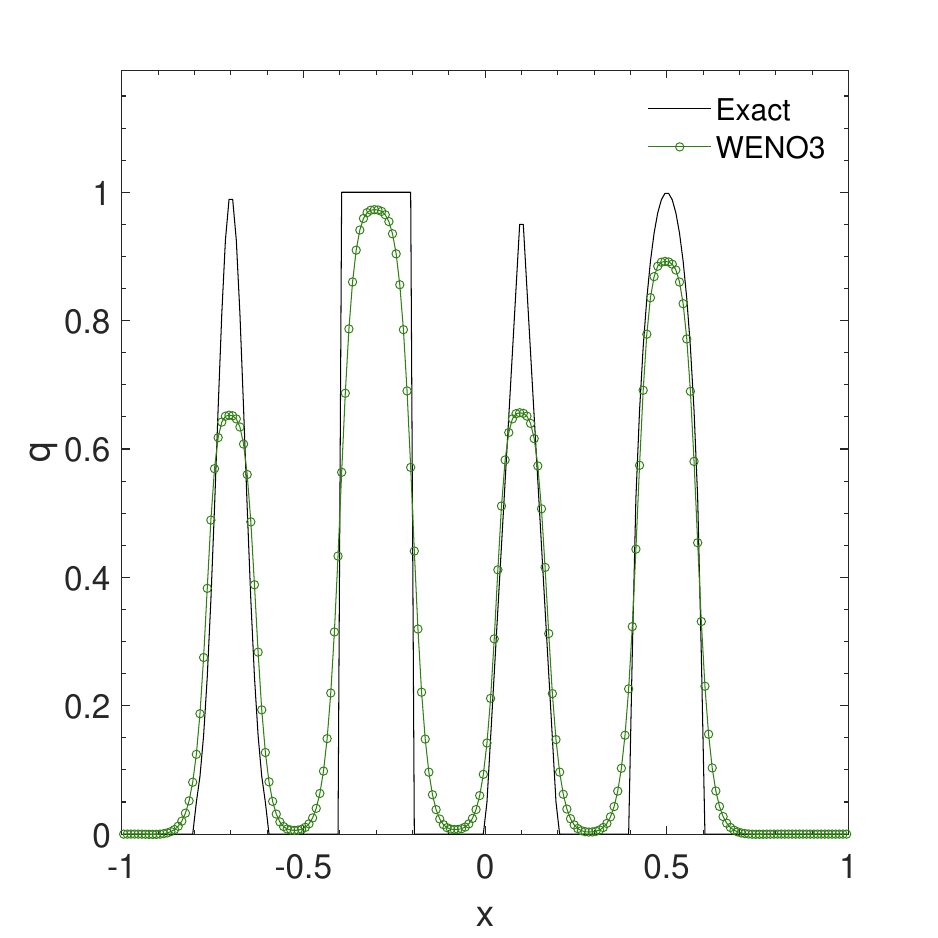}}
\subfigure[One-step BVD]
{\includegraphics[width=0.32\textwidth]{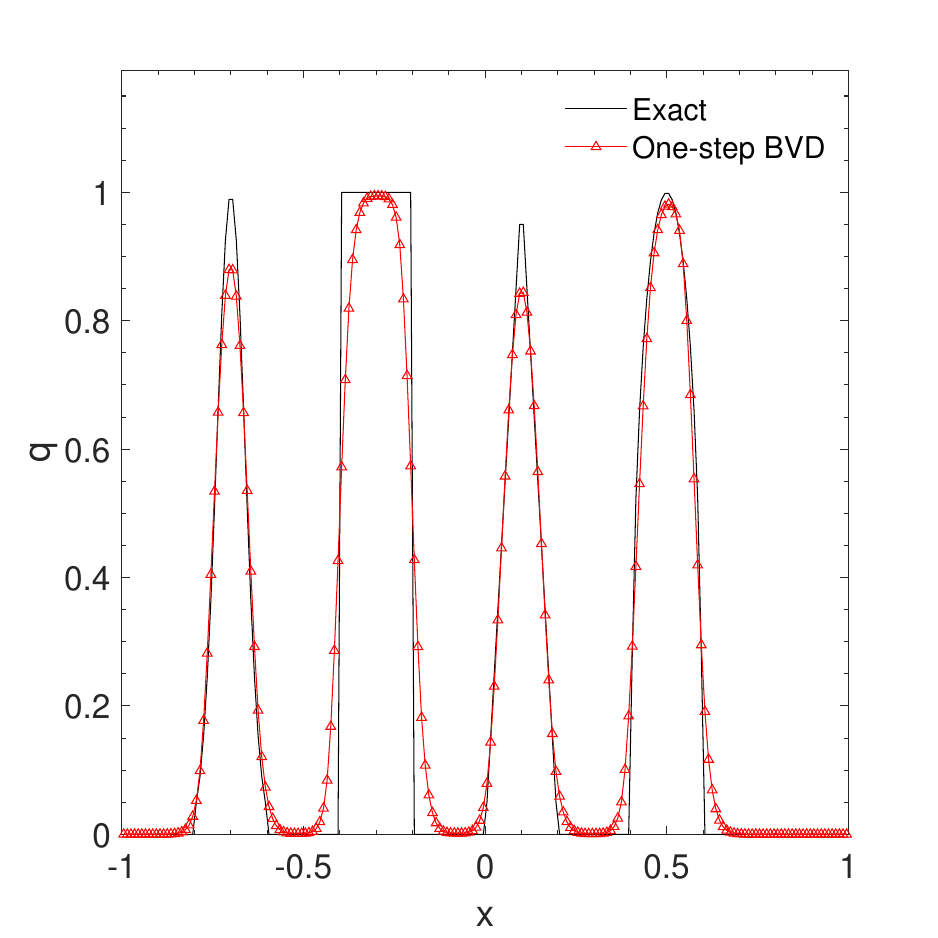}}
\subfigure[Two-step BVD]
{\includegraphics[width=0.32\textwidth]{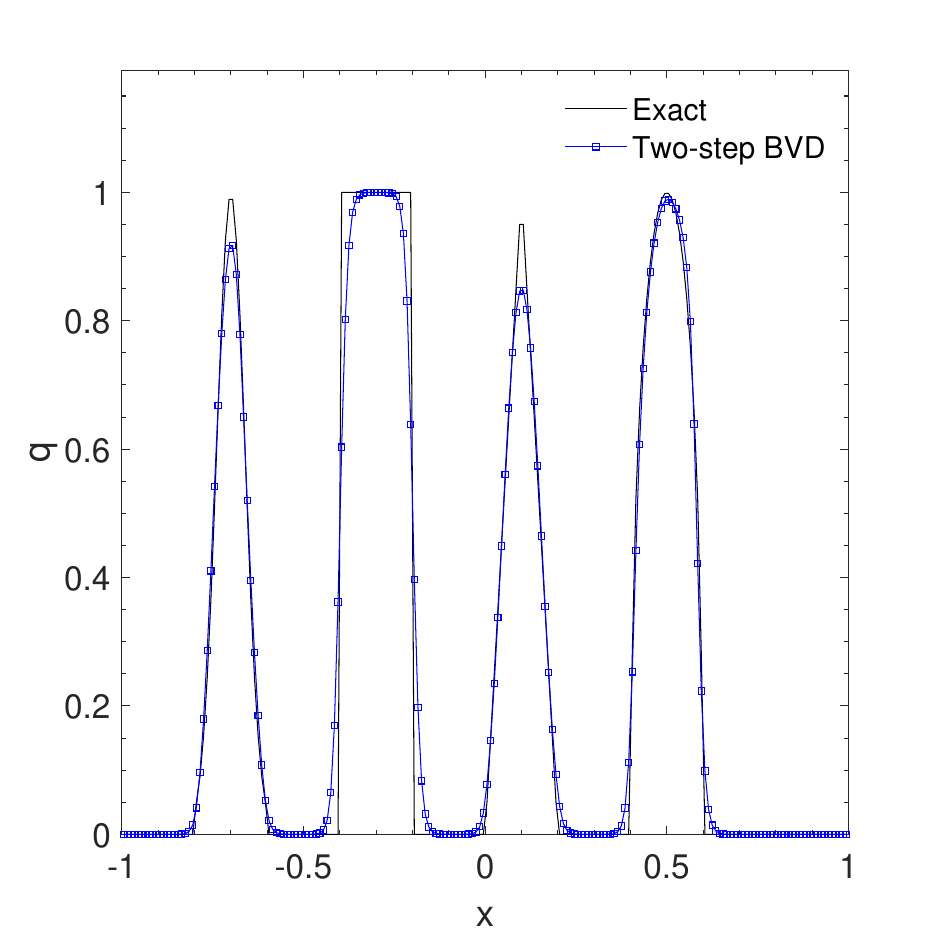}}
\caption{Numerical results of advection of Jiang and Shu's complex wave at $t=2$ (N=200) by different schemes.}
\label{JiangAndShuPlot}
\end{figure}

\begin{figure}[htbp]
\centering
\subfigure[WENO3]
{\includegraphics[width=0.32\textwidth]{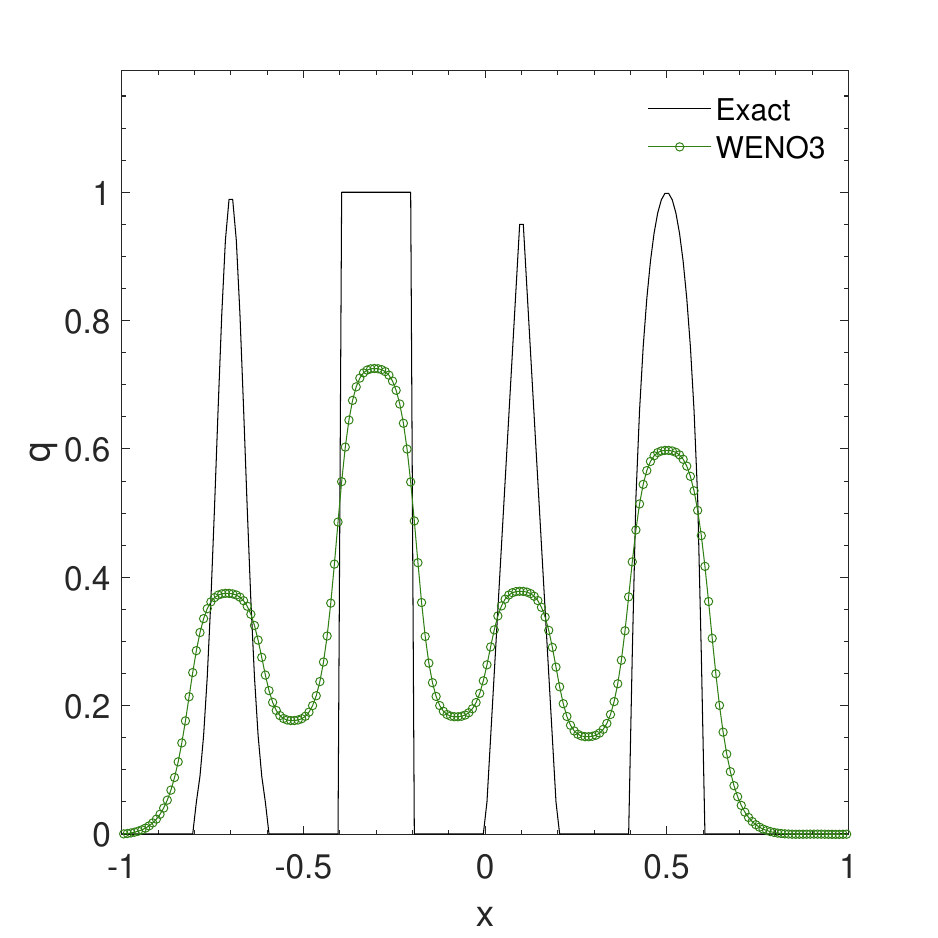}}
\subfigure[One-step BVD]
{\includegraphics[width=0.32\textwidth]{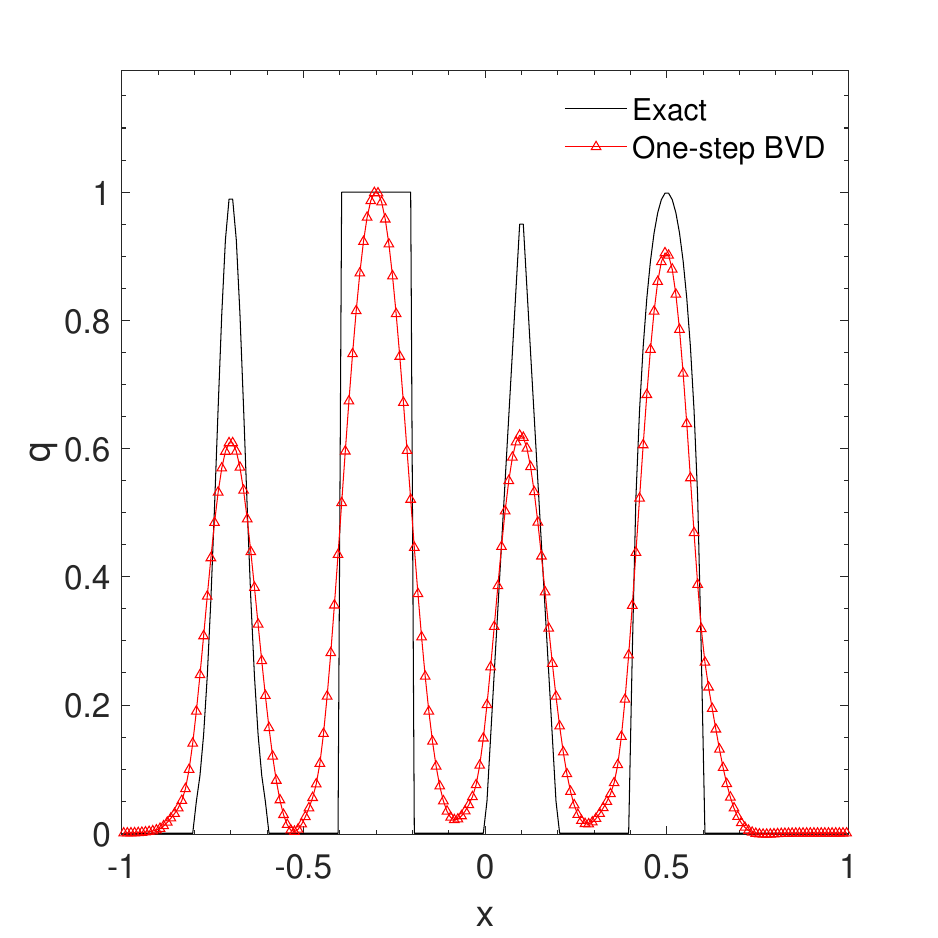}}
\subfigure[Two-step BVD]
{\includegraphics[width=0.32\textwidth]{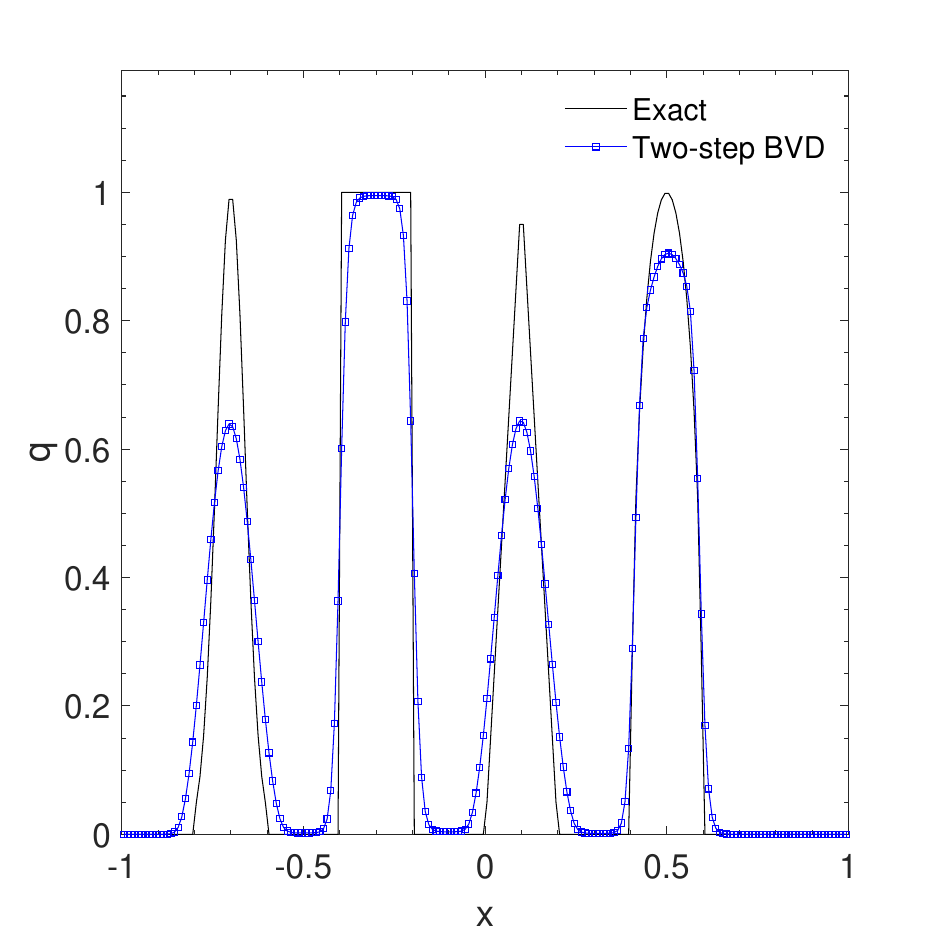}}
\caption{Same as Fig. \ref{JiangAndShuPlot}, but for numerical results at $t=20$.}
\label{JiangAndShuPlot20}
\end{figure}

The numerical results at $t=2$ with 200 computational cells by three different schemes are shown in Fig. \ref{JiangAndShuPlot}. It can be found that the proposed BVD schemes produce more accurate results and better preserve the extreme values in comparison with the WENO3 scheme. By introducing the second step in BVD algorithm, an anti-diffusion correction further improves the accuracy of the BVD scheme. In order to evaluate the performance of the schemes for long-time simulations, we also show the numerical results at $t=20$ (after integrating the advection models for 10 complete periods) in Fig. \ref{JiangAndShuPlot20}. The difference between the numerical results by three schemes become obviously larger. Two BVD schemes are much more accurate than WENO3 scheme.

\subsection{Test cases for 1D MHD equations}

Two test cases for 1D MHD flows are carried out in this part and for the sake of brevity we only give the numerical solutions by two-step BVD and WENO3 schemes for comparison. But it is should be pointed out that one-step BVD also has advantage in the solution fidelity in comparison with WENO3 scheme and more efficient than two-step algorithm. It is worth investigations in practical applications to choose one or complete two-step BVD reconstruction considering the overall performance of the computational accuracy and efficiency.

\subsubsection{Brio-Wu shock tube problem}\label{Brio-Wu shock tube problem}

The Brio-Wu problem was extended from Sod shock tube problem of Euler equations \citep{Brio_1988}. It is widely used as a benchmark test for one-dimensional ideal MHD equations. The initial conditions are specified as
\begin{equation}
\left(\rho, u, v, w, B_x, B_y, B_z, p\right) =
\left\{\begin{array}{ll}
\left(1    , 0, 0, 0, 0.75,  1, 0,   1 \right) & \mathrm{if}\ x<0    \\
\left(0.125, 0, 0, 0, 0.75, -1, 0, 0.1 \right) & \mathrm{otherwise}    \\
\end{array}\right.
\end{equation}
and $\gamma=2$ is used.

\begin{figure}[htbp]
\centering
\subfigure[Density]	
{\includegraphics[width=0.45\textwidth]{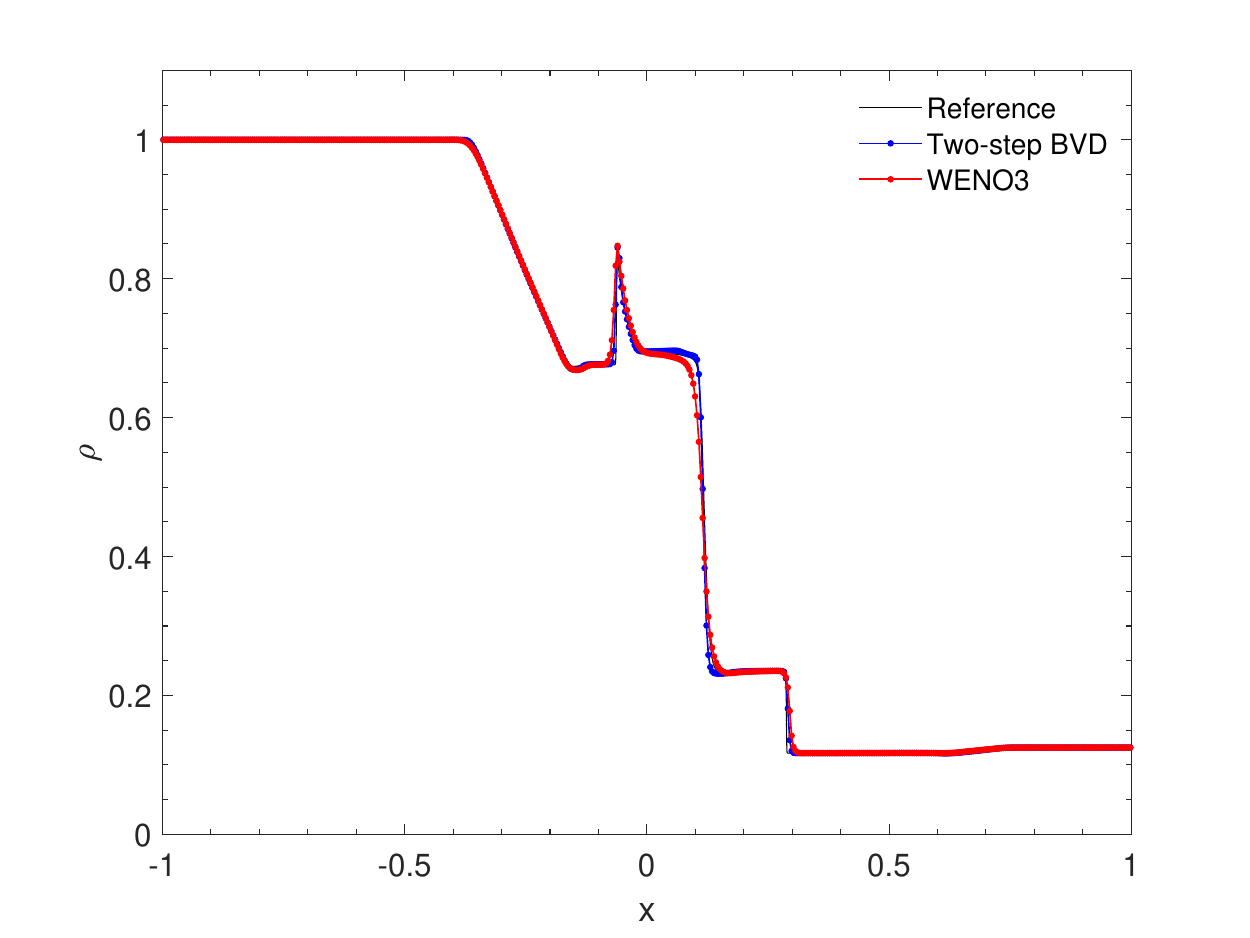}}
\subfigure[Pressure]	
{\includegraphics[width=0.45\textwidth]{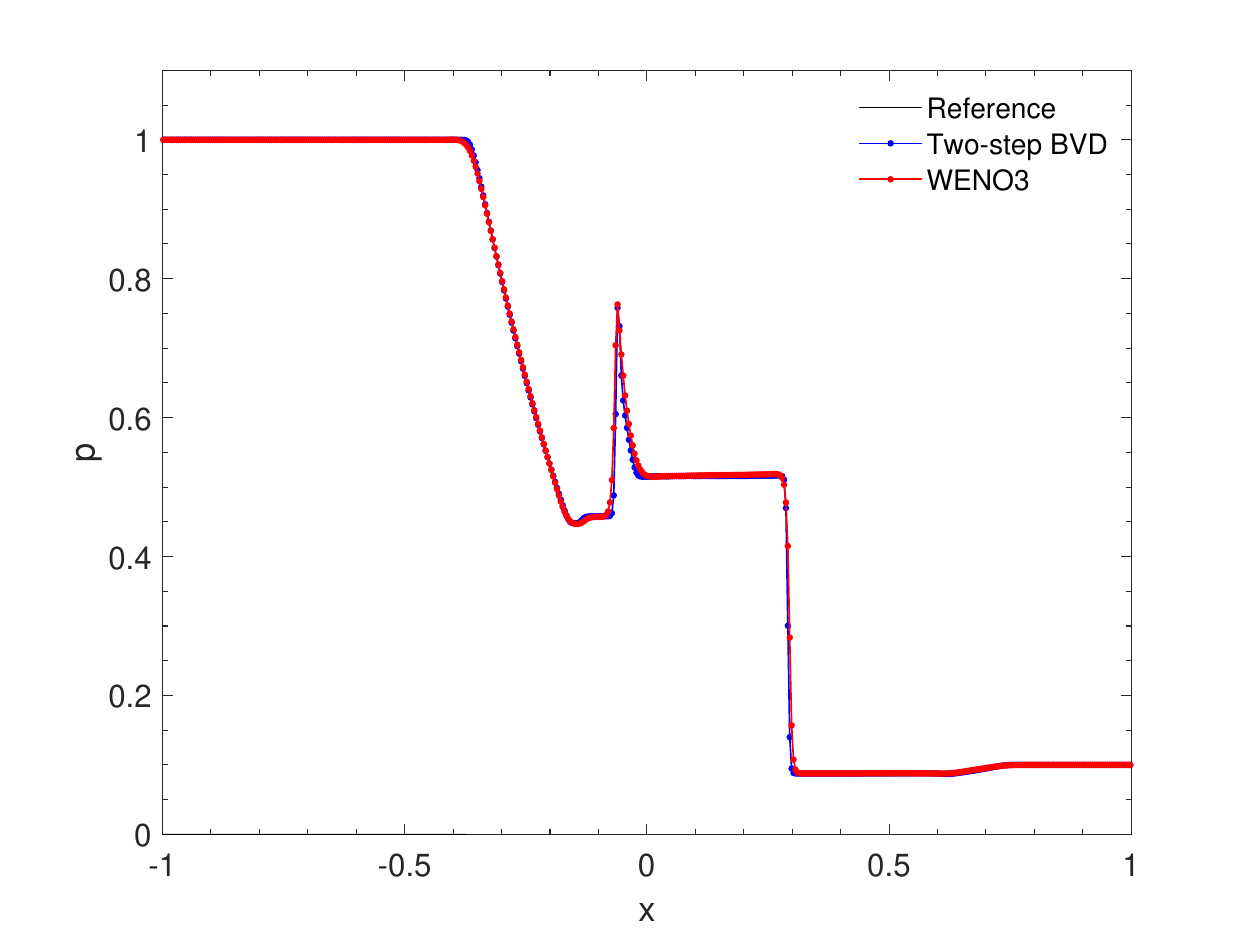}}
\subfigure[Velocity (x component)]	
{\includegraphics[width=0.45\textwidth]{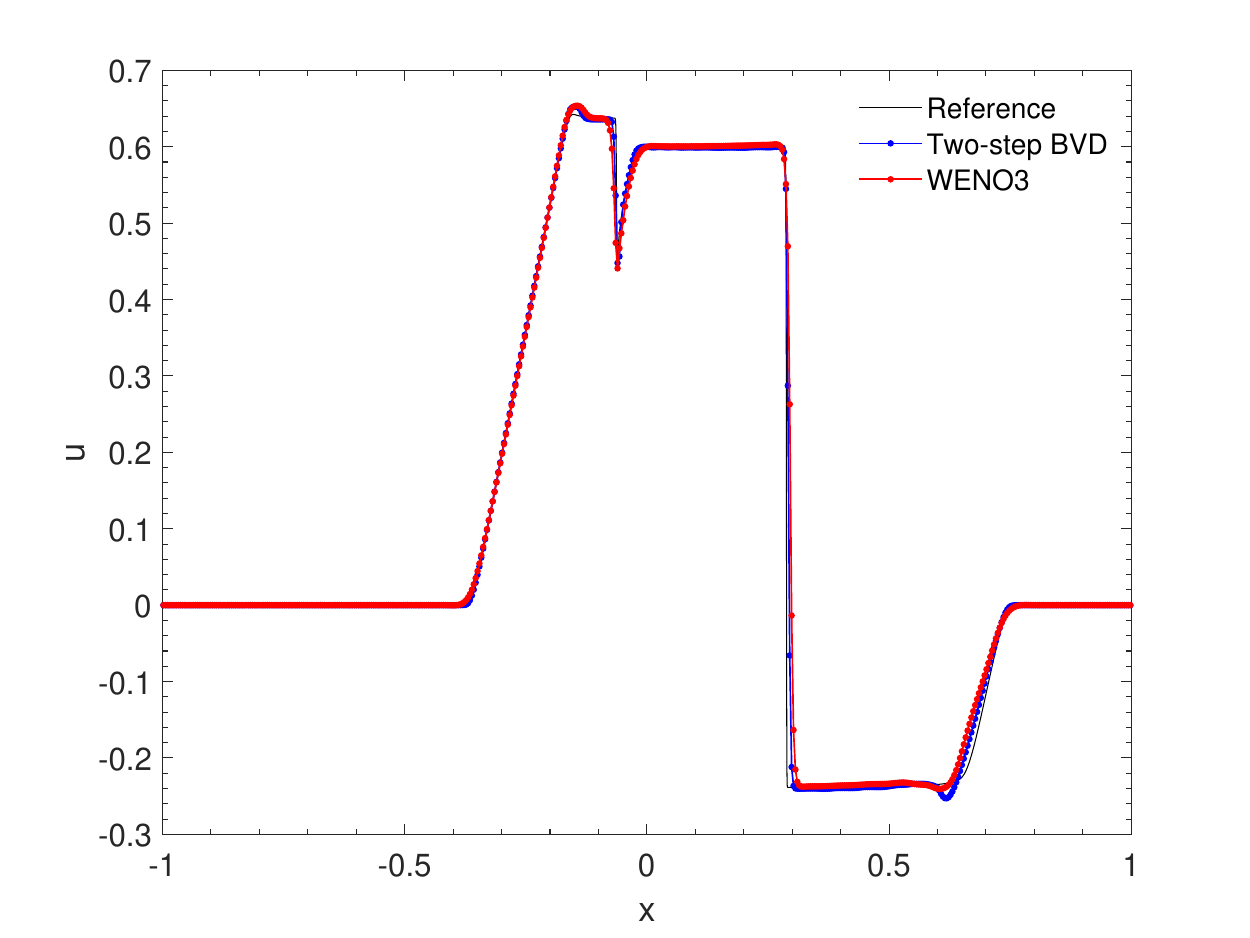}}
\subfigure[Magnetic field (y component)]	
{\includegraphics[width=0.45\textwidth]{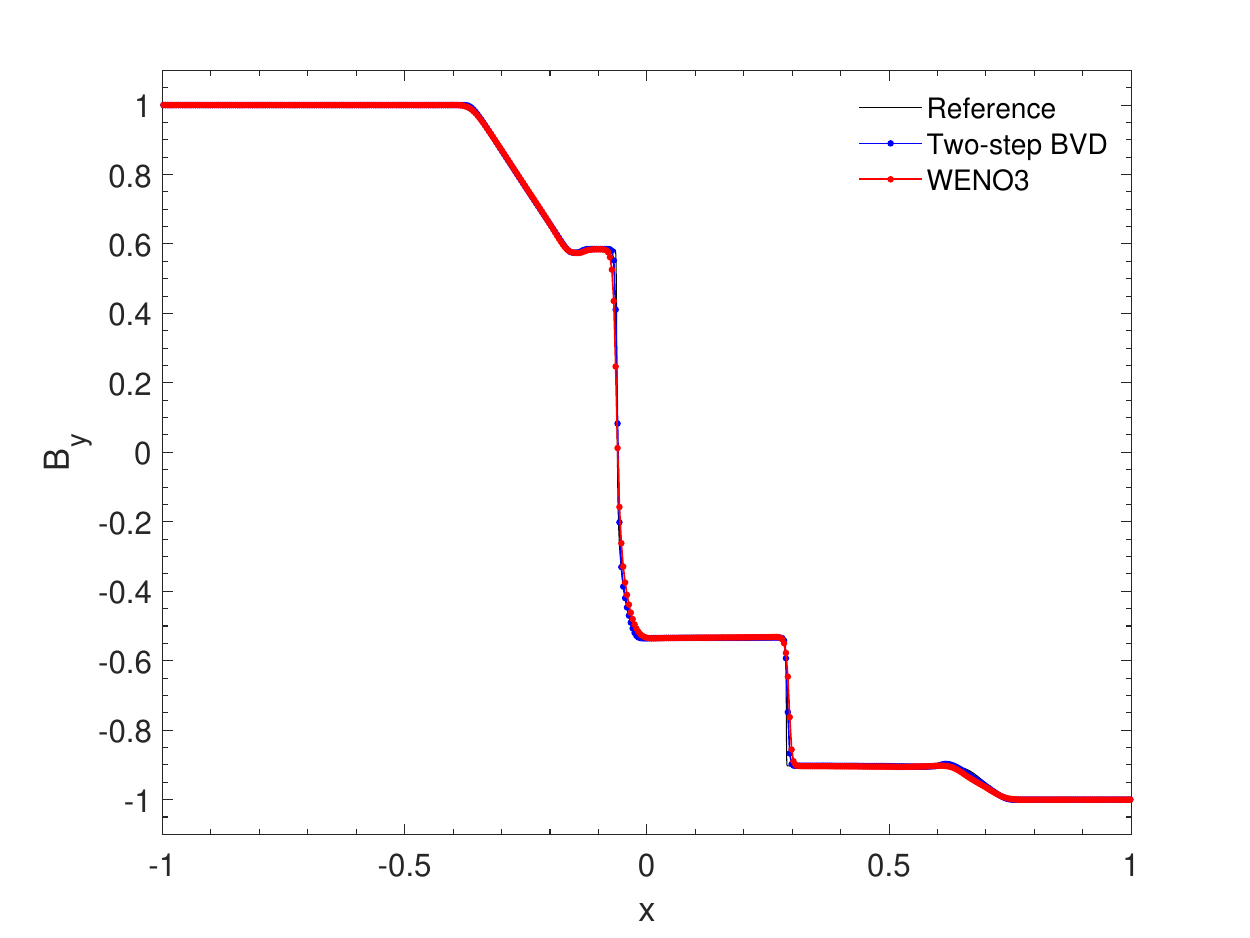}}
\caption{Numerical results of Brio-Wu problem at $t=0.2$ (N=512) by two-step BVD and WENO3 schemes. }
\label{Brio-Wu}
\end{figure}

The numerical results with 512 cells at $t=0.2$ are shown in Fig. \ref{Brio-Wu} for the density, pressure, x component of velocity and y component of magnetic field by two-step BVD and WENO3 schemes. The reference solution is computed using a $1^{st}$-order upwind scheme with a very high grid resolution. Both schemes accurately capture the wave structures of MHD flows and the results agree well with the reference one. Around the discontinuities, two-step BVD scheme produces more sharp results, which reveals it is less dissipative in comparison with WENO3 scheme.

\subsubsection{High Mach number shock tube problem}
The second test case is carried out to demonstrate the robustness of the proposed scheme for a high Mach number MHD flow. The Mach number corresponding to the right-moving shock wave is $15.5$. The initial conditions \citep{Brio_1988,Jiang_1999} are specified as
\begin{equation}
\left(\rho, u, v, w, B_x, B_y, B_z, p\right) =
\left\{\begin{array}{ll}
\left(1,0,0,0,0,1,0,1000 \right) & \mathrm{if}\ x<0    \\
\left(0.125,0,0,0,0,-1,0,0.1 \right) & \mathrm{otherwise}    \\
\end{array}\right.
\end{equation}
with $\gamma=2$.

\begin{figure}[htbp]
\centering
\subfigure[Density]	
{\includegraphics[width=0.45\textwidth]{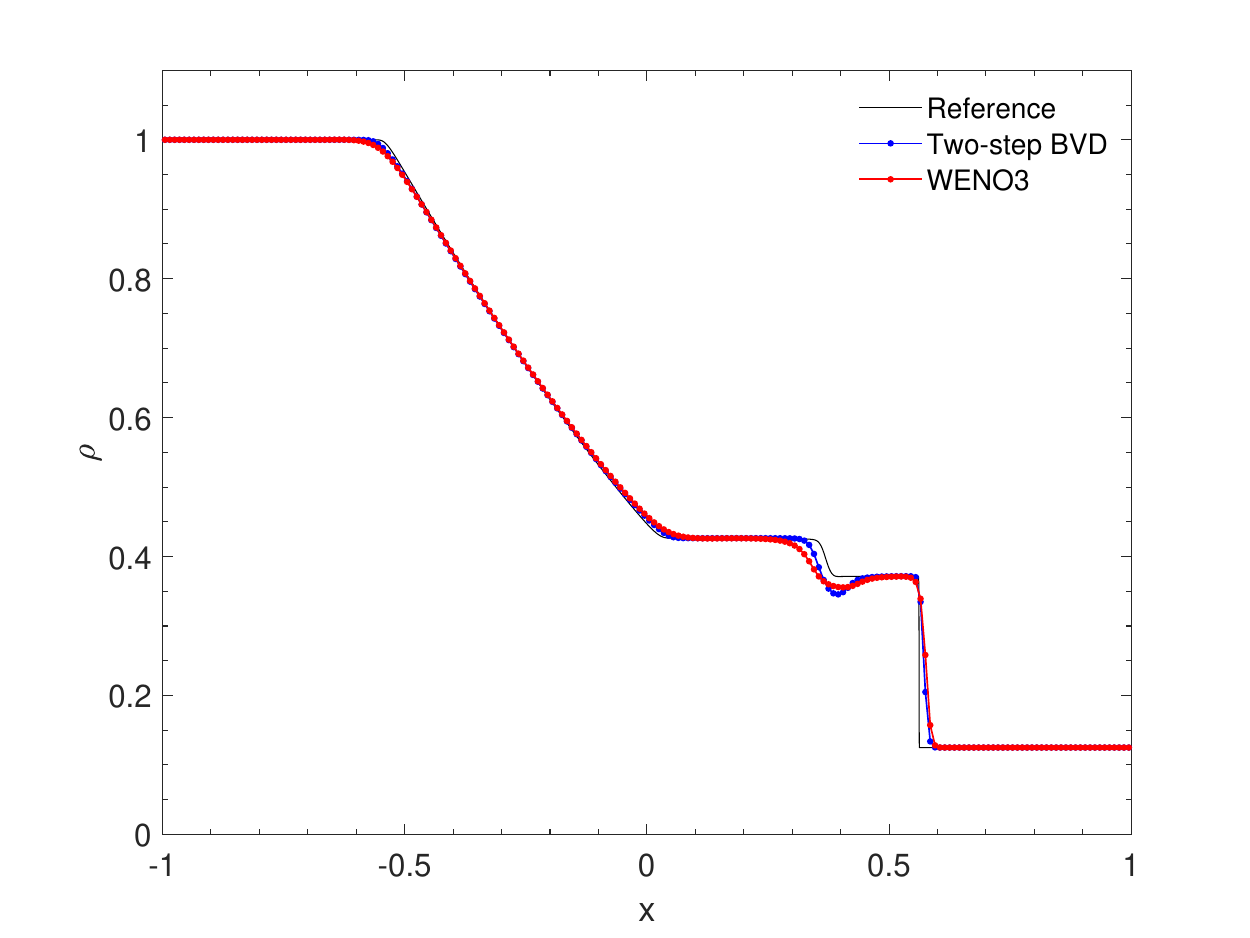}}
\subfigure[Pressure]	
{\includegraphics[width=0.45\textwidth]{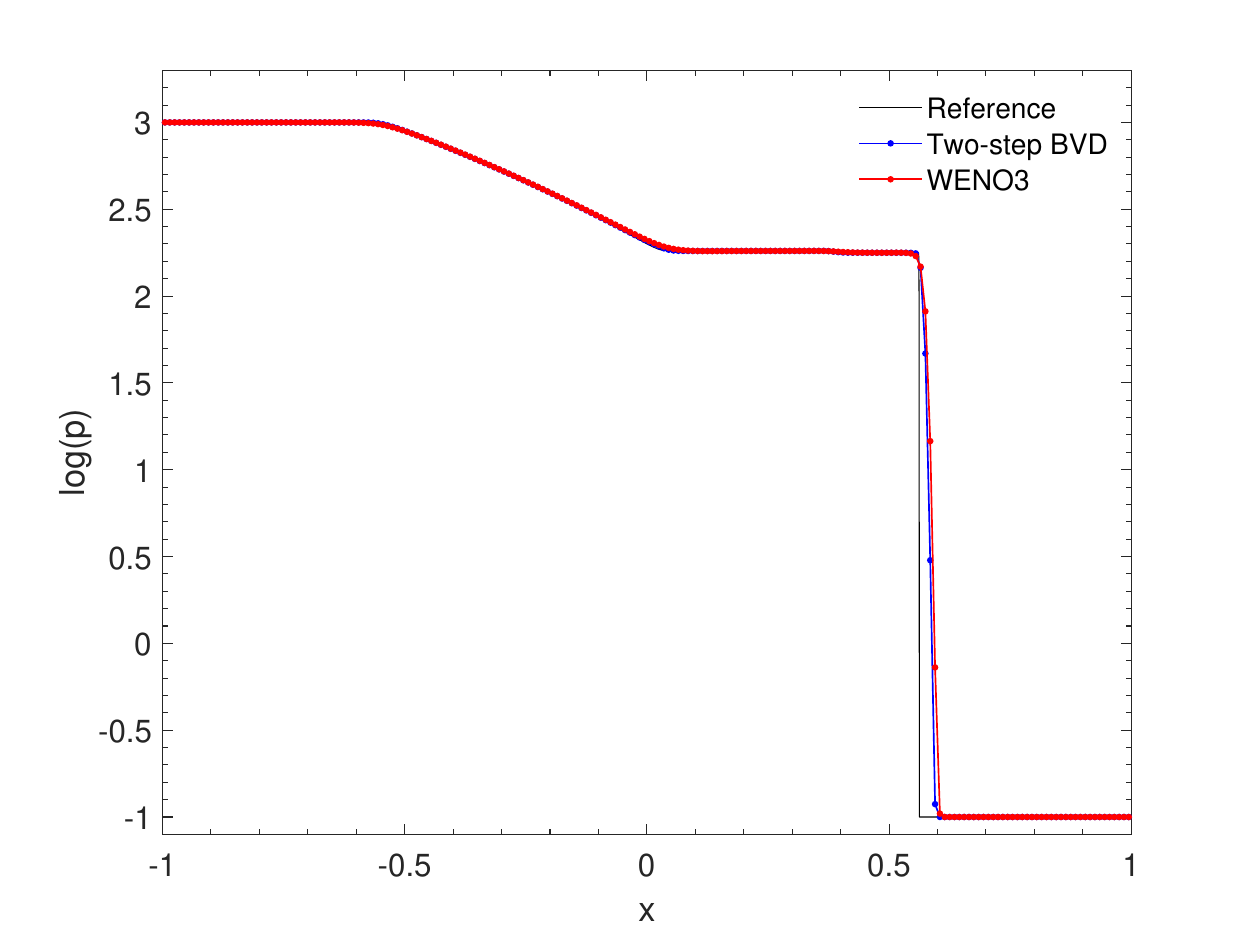}}
\subfigure[Velocity (x component)]	
{\includegraphics[width=0.45\textwidth]{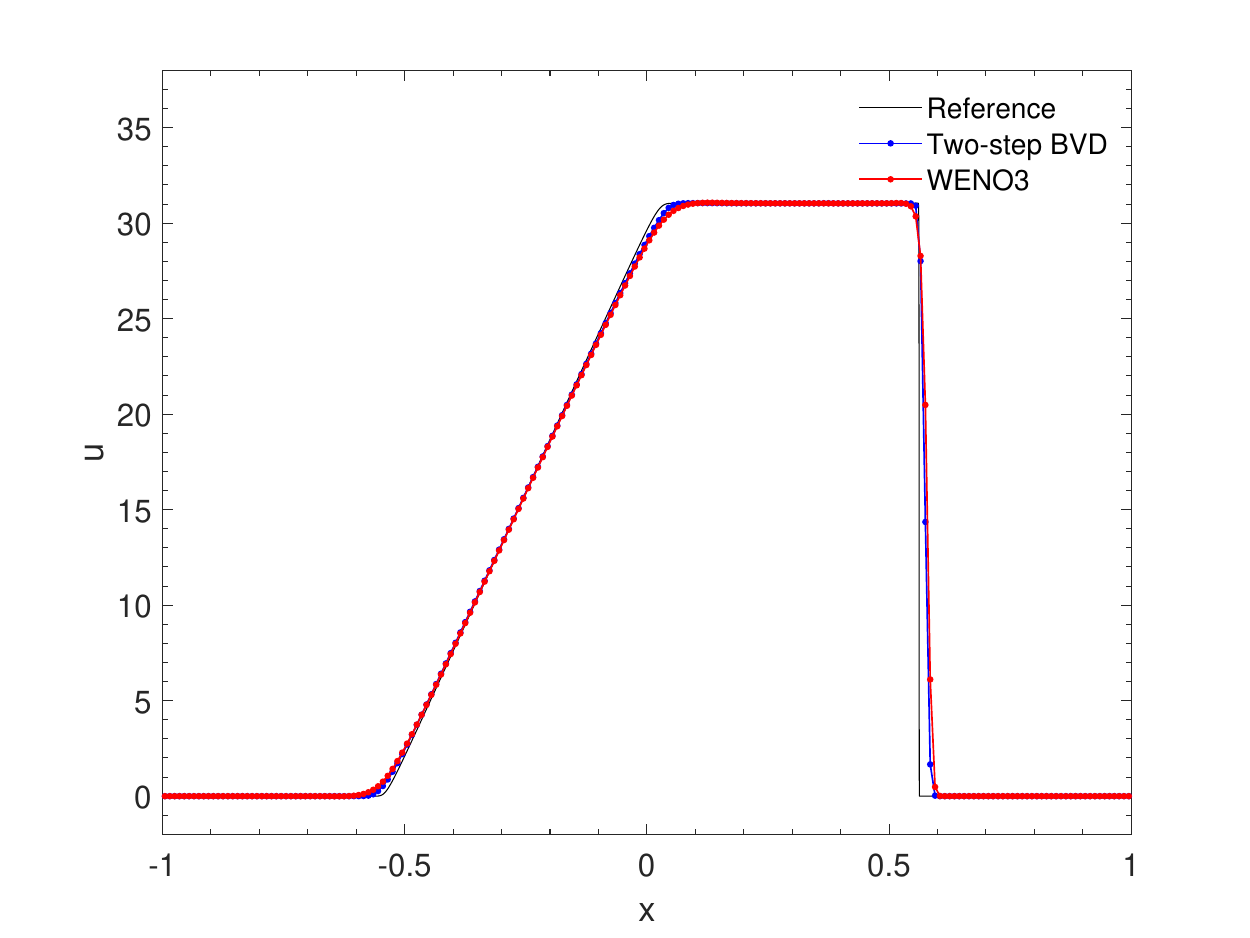}}
\subfigure[Magnetic field (y component)]	
{\includegraphics[width=0.45\textwidth]{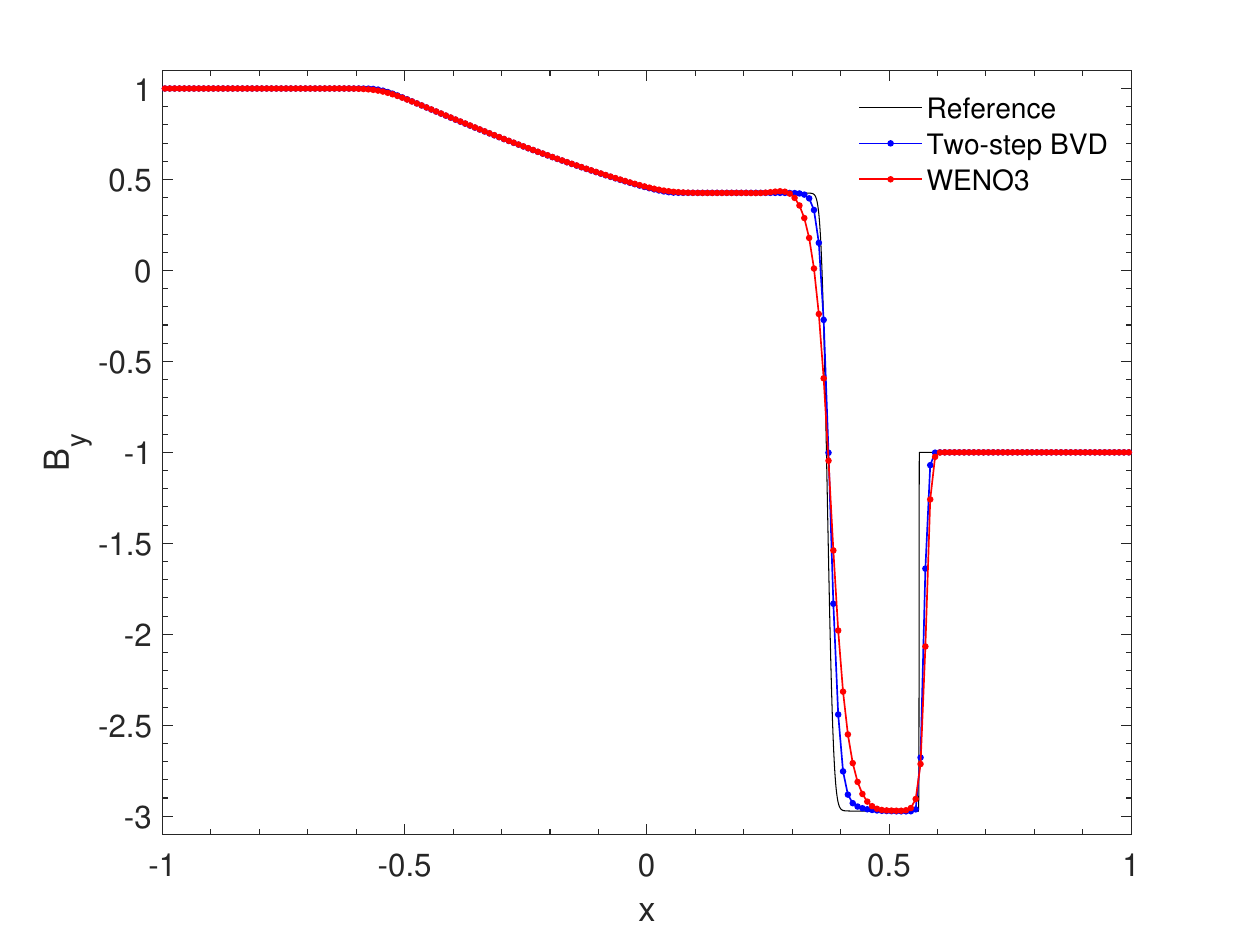}}
\caption{Numerical results of high Mach number shock tube problem at $t=0.012$ (N=200) by two-step BVD and WENO3 scheme.}
\label{highmach}
\end{figure}

The numerical result with $200$ cells at $t=0.012$ is shown in Fig. \ref{highmach}. Both schemes can stably calculate this test without generates nonphysical negative pressure. The numerical results agree well with the reference one, except a slight deviation found around $x=0.4$ for the density. Similar with previous test, two-step BVD scheme produces more accurate solutions of discontinuities, especially for the numerical results of the y component of magnetic field.

\subsection{Test cases for 2D MHD equations}

\subsubsection{2D circularly polarized Alfv\'{e}n wave}\label{2D circularly polarized Alfven wave}

The 2D circularly polarized Alfv\'{e}n wave problem \citep{Toth_2000,Liu_2021} is used to check the convergence rates of the proposed schemes by calculating a smooth MHD flow. We specify the propagation angle of the Alfv\'{e}n plane wave with respect to x axis as $\alpha=\frac{\pi}{4}$. The computational domain is $\left(x,y\right) \in \left[0,\frac{1}{\cos \alpha}\right]\times\left[0,\frac{1}{\sin \alpha}\right]$. The initial conditions are
\begin{equation}\label{Init-Alfven}
\rho =1, \ v_{\parallel}=0, \ v_{\perp}=0.1 \sin \left(2\pi \beta\right), \ w=0.1 \cos \left(2\pi \beta\right), \ p=0.1, \ B_{\parallel}=1, \ B_{\perp}=v_{\perp}, \ {\rm and}\ B_z=w,
\end{equation}
where $\beta=x\cos \alpha + y \sin \alpha $ and the subscripts $\parallel$ and $\perp$ denote the directions parallel and perpendicular to the wave propagation respectively.

In this test, we set $\gamma=\frac{5}{3}$. Because speed of the Alfv\'{e}n wave is $c_a=\frac{B_{\parallel}}{\sqrt{\rho}}=1$, the flow field returns to its initial state at time $t=1,2,3,...$. We calculate this test on a series of refining uniform grids and the time step is chosen as $\Delta t=\frac{0.4}{N}$ where $N$ is the cell number in either direction \citep{Toth_2000,Liu_2021}. In this test, the non-divergent correction is omitted as suggested in \cite{Shen_2012}, since the CT algorithm used in this study is of $2^{nd}$-order accuracy.

Table \ref{2D_MHD_accuracy} shows the $l_1$ errors and orders of two components of velocity and magnetic fields at $t=2$. Two BVD schemes achieve the identical accuracy as the linear upwind scheme, except two-step BVD scheme is found to be more accurate on the coarsest grid. WENO3 scheme produces larger errors and also shows slower convergence rates. At the finest grid, the errors of WENO3 scheme are more than one order of magnitude larger than those of other three schemes.

\subsubsection{Orszag-Tang turbulence problem}

Orszag-Tang turbulence problem has widely been used to check the performance of numerical models for MHD equations \citep{Orszag_1979,Toth_2000,Jiang_1999,Shen_2012,Liu_2021}. Its solution reflects representative characteristics of MHD turbulence, involving the complex interactions between shock waves and complex vortex structures.

Following \cite{Shen_2012}, the initial conditions are specified as
\begin{equation}\label{Init-OZ}
\left(\rho,u,v,w,B_x,B_y,B_z,p\right)=\left(\gamma^2,-\sin y,\sin x,0,-\sin y,\sin 2x,0,\gamma\right),
\end{equation}
where $\gamma=5/3$, the computational domain is $\left(x,y\right) \in \left[0,2\pi\right] \times \left[0,2\pi\right]$ and the periodic boundary conditions are used in this test.

\begin{figure}[htbp]
\centering
\subfigure[density]
{\includegraphics[width=0.45\textwidth]{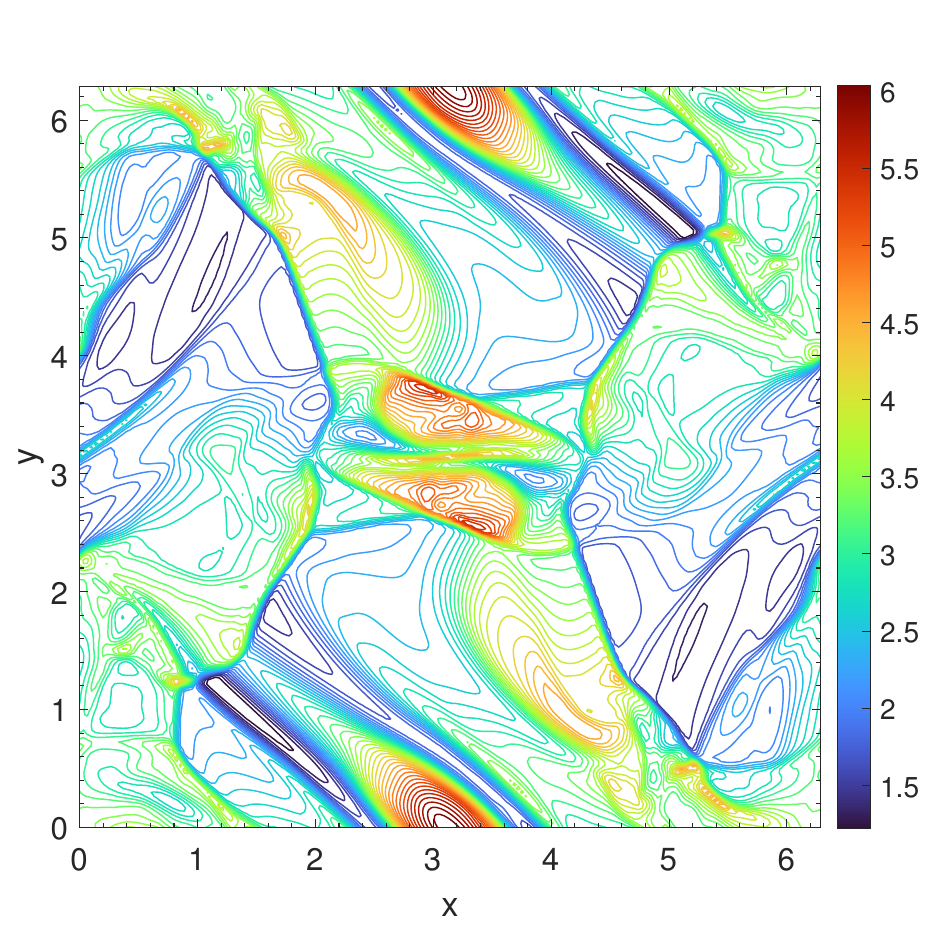}}
\subfigure[Pressure]
{\includegraphics[width=0.45\textwidth]{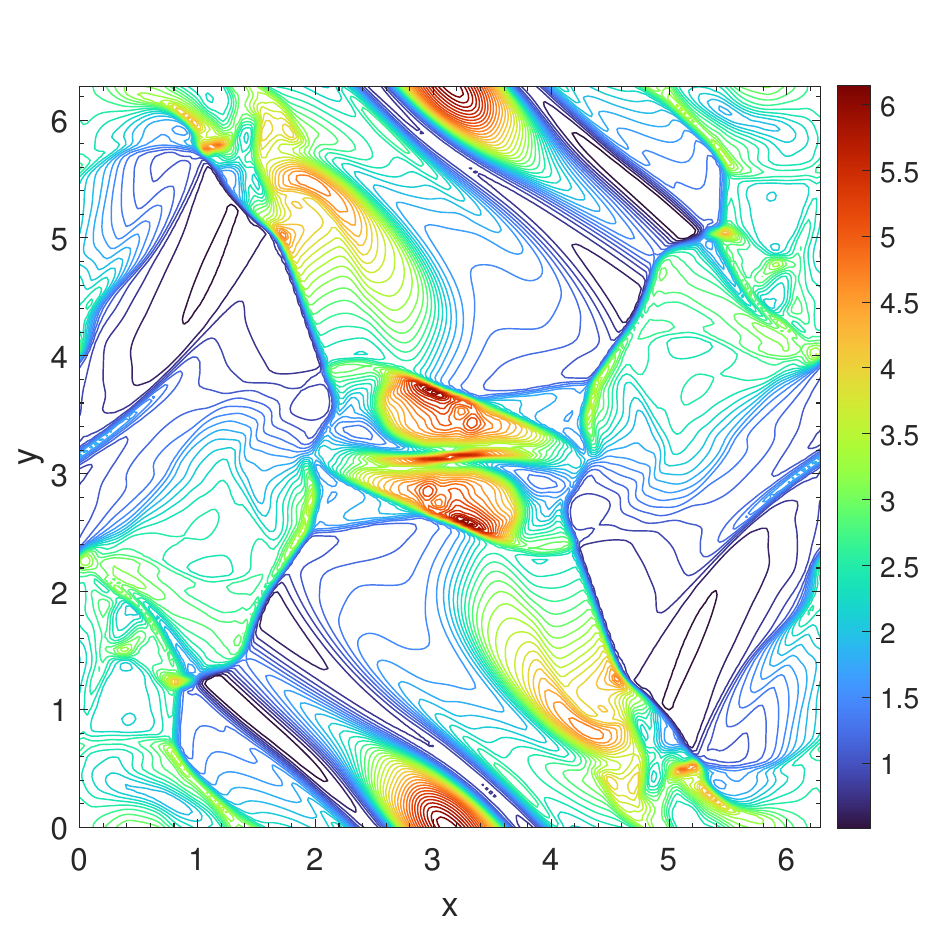}}\\
\subfigure[Magnetic pressure\label{p_P2T2}]
{\includegraphics[width=0.45\textwidth]{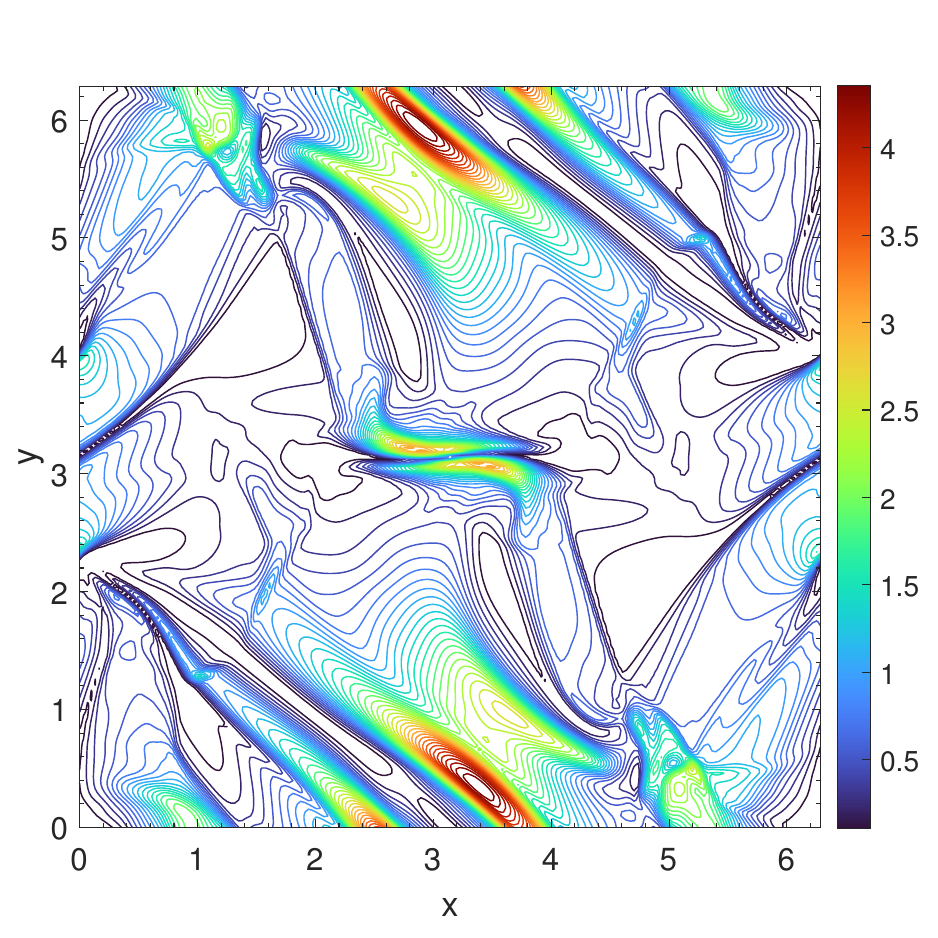}}
\subfigure[Specific kinetic energy]
{\includegraphics[width=0.45\textwidth]{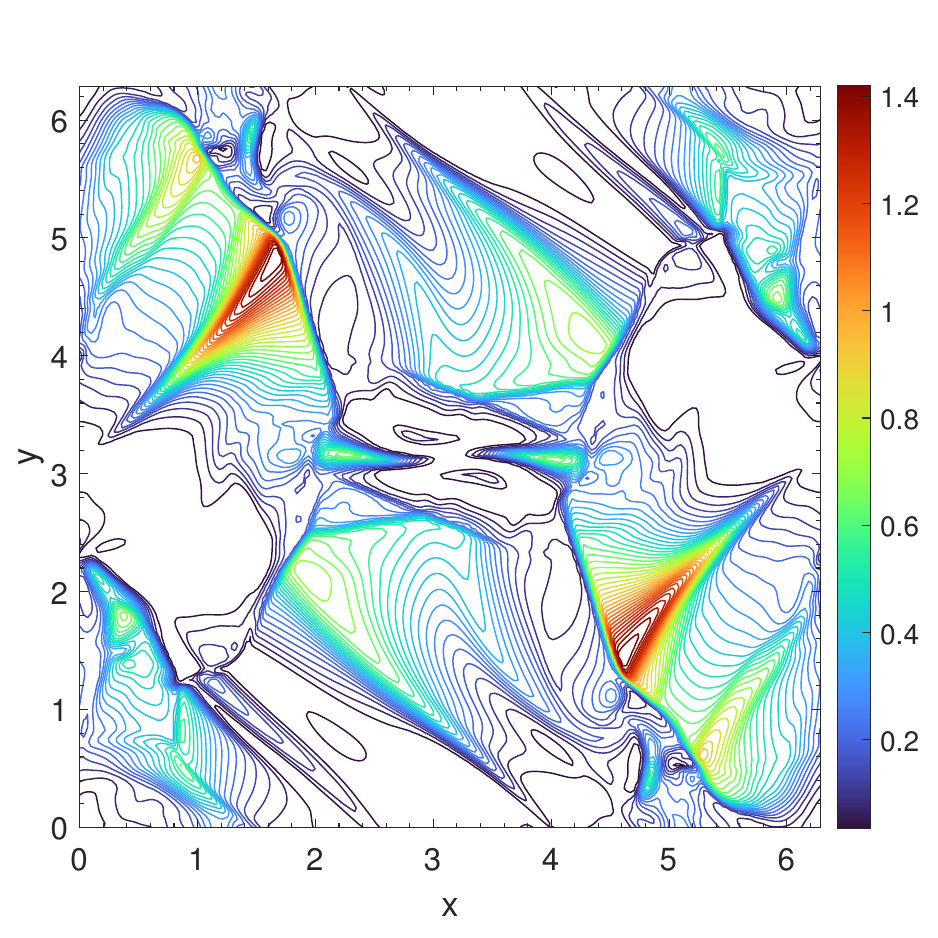}}\\
\caption{Contour plots of numerical results of Orszag-Tang turbulence problem at $t=3$ (N=192) by two-step BVD scheme.}\label{OZ_contour}
\end{figure}

In Fig. \ref{OZ_contour}, we show the numerical results by two-step BVD scheme using $192 \times 192$ cells at $t=3$ for the density, pressure, magnetic pressure ($\frac{1}{2}\boldsymbol{B}\cdot\boldsymbol{B}$) and specific kinetic energy ($\frac{1}{2}\boldsymbol{v}\cdot\boldsymbol{v}$). The solutions of Orszag-Tang turbulence problem contain complex interactions between shocks and multi-scale vortex structures. Our results are consistent with reference ones given in \cite{Orszag_1979,Toth_2000,Jiang_1999,Shen_2012,Liu_2021}.
The proposed BVD scheme has ability to reproduce the complex MHD flows with high fidelity.

\begin{figure}[htbp]
\centering
\includegraphics[width=0.6\textwidth]{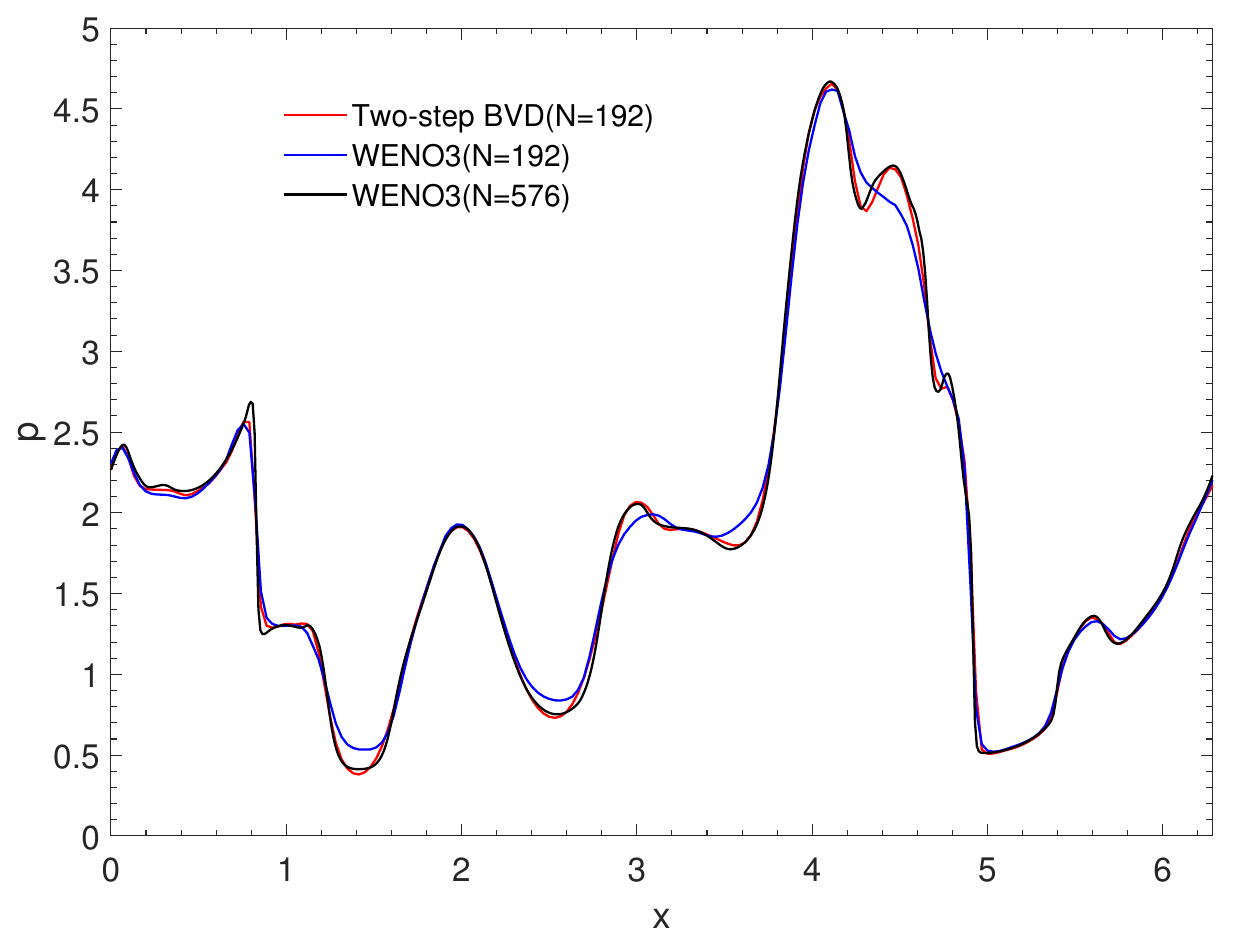}
\caption{Cross-section plots of numerical results of pressure along line $y=1$.}\label{OZ_p_slice}
\end{figure}

To quantitatively evaluate the performance of the proposed scheme,
we give a cross-section plot of numerical results of the pressure along line $y=1$ in Fig. \ref{OZ_p_slice}. The shown results are calculated by the two-step BVD scheme with 192 cells, WENO3 scheme with 192 and 576 cells. Using the same grid resolution, BVD schemes can reproduce more accurate solution. Increasing the grid resolution by 3 times, the results of WENO3 scheme are obviously improved and more details of MHD flows can be observed. The proposed BVD scheme give the solution of the similar quality on much coarser grid in comparison with WENO3 scheme.

\subsubsection{Rotor problem}\label{RotorProblem}

A rotor problem introduced by \cite{Balsara_1999} is carried out for the numerical simulation of the propagation of torsional Alfv\'{e}n wave. The initial conditions are given by
\begin{eqnarray}\label{Init-RotorProblem}
\rho &=&1+9f(r), \\
\left( u, v, w\right) &=&\left\{
\begin{array}{ll}
\left(-\frac{2f(r)y}{0.1},  \frac{2f(r)x}{0.1}, 0 \right) & \mathrm{if}\ r <0.1    \\
\left(-\frac{2f(r)y}{ r },  \frac{2f(r)x}{ r }, 0 \right) & \mathrm{if}\ r \geqslant 0.1   \\
\end{array}\right., \\
\left( B_x, B_y, B_z, p\right) &=&\left( 5/\sqrt{4\pi} , 0,0,1 \right),
\end{eqnarray}
where $r=\sqrt{x^2+y^2}$,
\begin{equation}\label{f}
f(r)=\left\{
\begin{array}{ll}
1                                       & \mathrm{if}\ r <0.1 ,   \\
\frac{200}{3}\left( 0.115-r \right) \ \ & \mathrm{if}\ 0.1 \leqslant r \leqslant 0.115  \\
0                                       & \mathrm{if}\ r > 0.115 .\\
\end{array}\right.,
\end{equation}
the computational domain is $(x,y)\in[-0.5,0.5]\times [-0.5,0.5]$ and the ratio of specific heats $\gamma = 1.4$.

\begin{figure}[htbp]
\centering
\subfigure[Density]
{\includegraphics[width=0.45\textwidth]{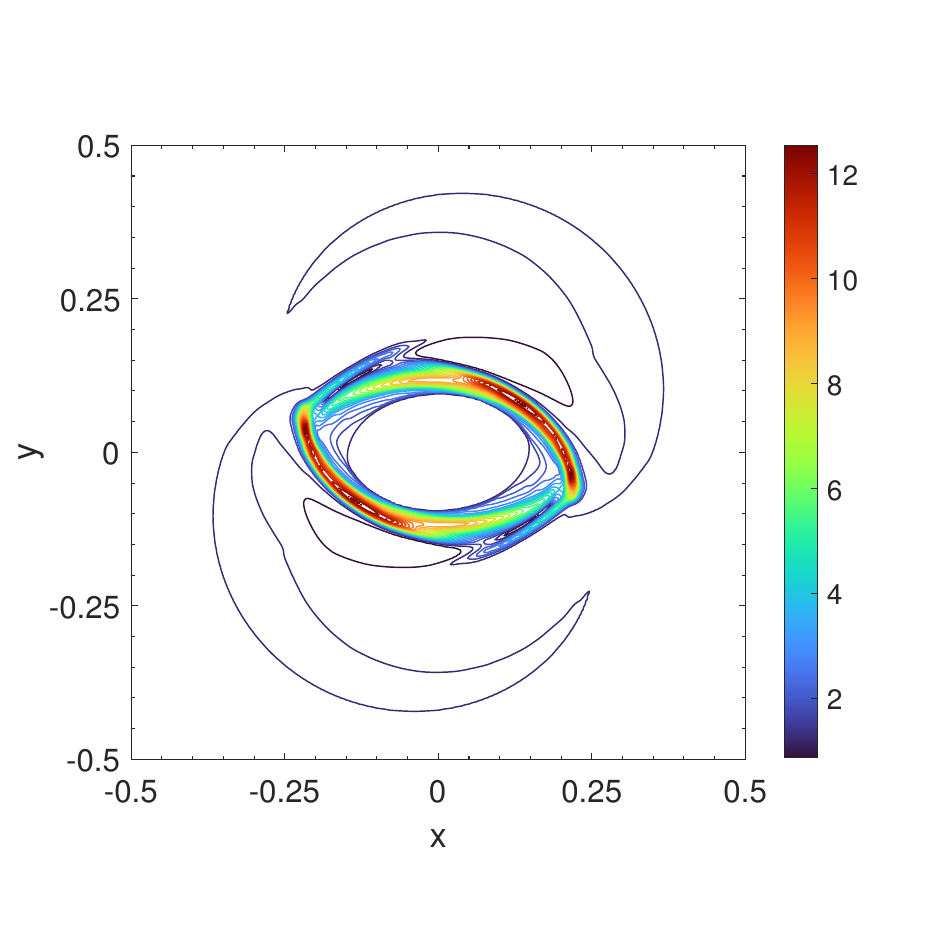}}
\subfigure[Pressure]
{\includegraphics[width=0.45\textwidth]{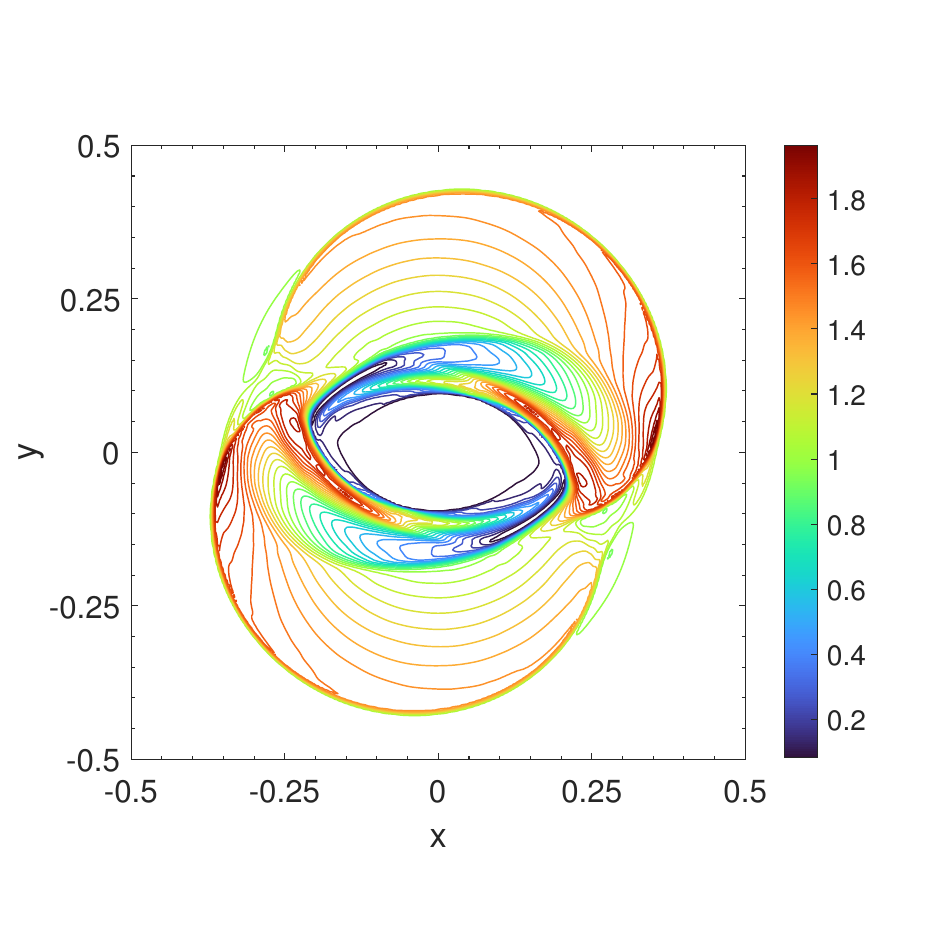}}\\
\subfigure[Magnetic pressure]
{\includegraphics[width=0.45\textwidth]{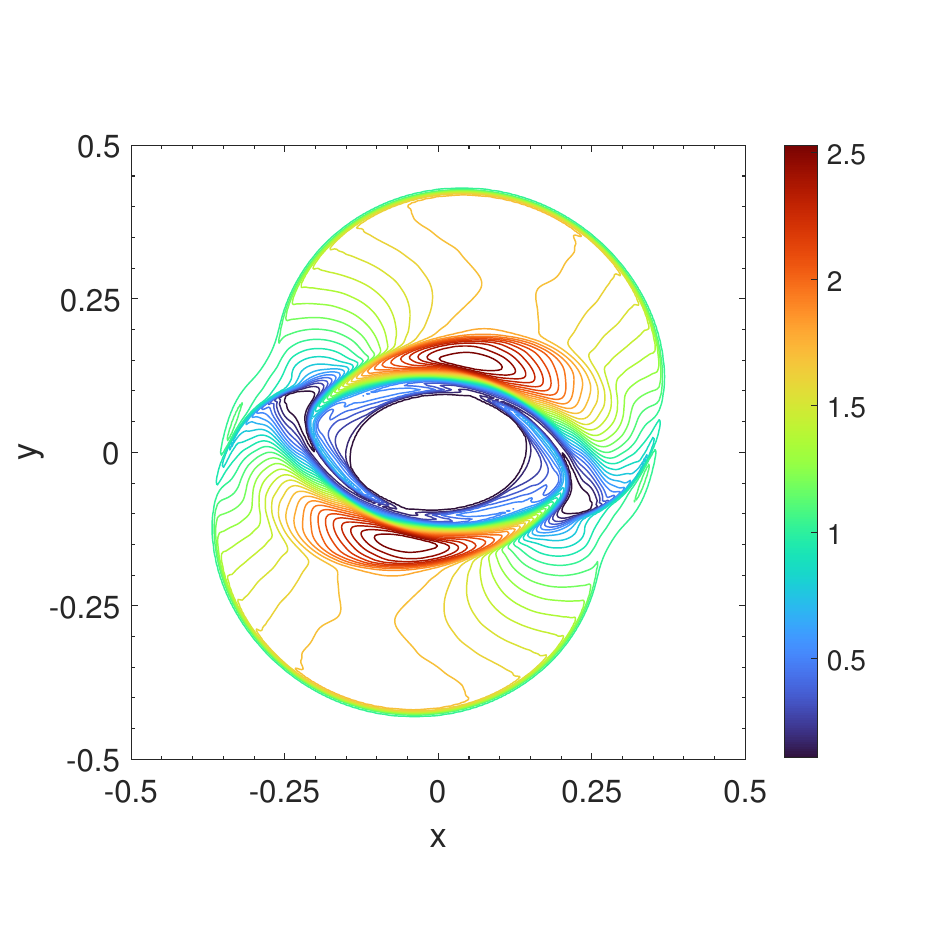}}
\subfigure[Ma]
{\includegraphics[width=0.45\textwidth]{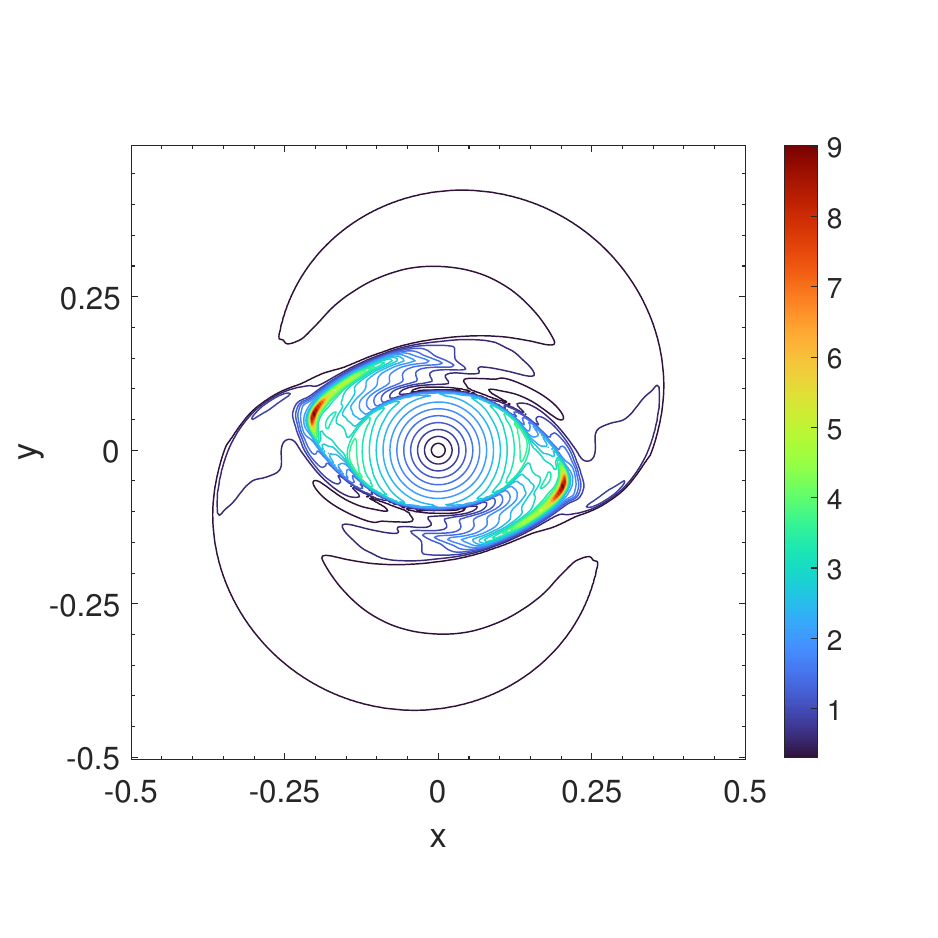}}\\
\caption{Contour plots of numerical results of 2D MHD rotor problem at $t=0.15$ (N=300) by two-step BVD scheme.}\label{RotorProblem_contour}
\end{figure}

\begin{figure}[htbp]
\centering
\includegraphics[width=0.6\textwidth]{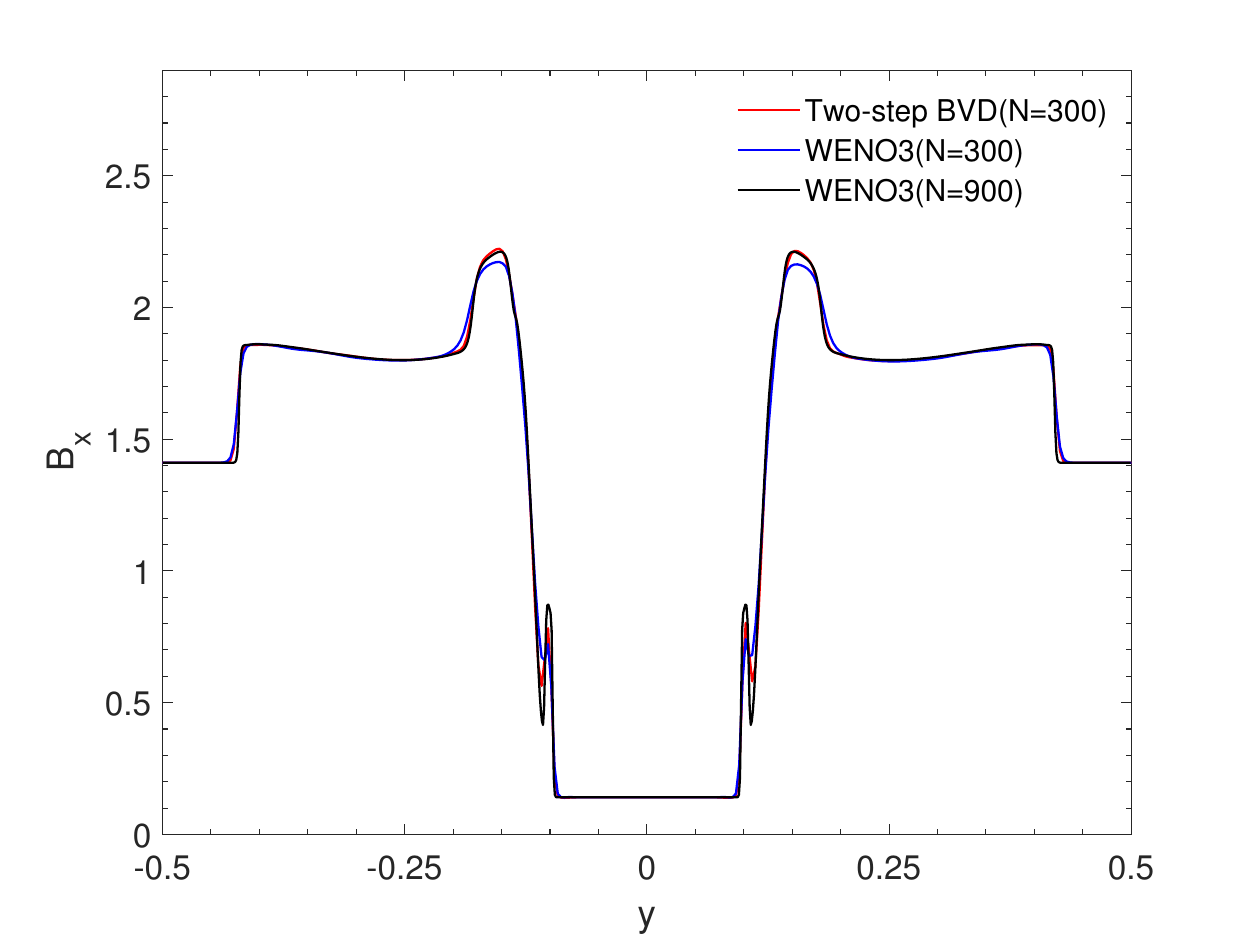}
\caption{Cross-section plots of numerical results of x component of magnetic field along line $x=0$.}\label{RotorProblem_profiles}
\end{figure}

The contour plots of numerical results of the density, pressure, magnetic pressure and Mach number on grid of $300 \times 300$ at $t =0.15$ by two-step BVD scheme are shown in Fig.\ref{RotorProblem_contour}. Again, the numerical results agree well with those presented in existing literatures \citep{Balsara_1999,Stone_2008,Shen_2012,Fu_2022}. Same as the previous test, we show a cross-section plot along line $x=0$ for results of x component of magnetic field by different schemes in Fig. \ref{RotorProblem_profiles}. The WENO3 scheme are used to calculate this test on two different grids $300\times 300$ and $900\times 900$ for comparison. The proposed BVD scheme produces better results in comparison with the WENO3 scheme on the same grid and shows much smaller deviations to the numerical result on a fine grid using one-ninth size of computational cells. Also, we found the solution of MHD flow involves the structures over a wide range of spatial scales, e.g. the fluctuations around $x=\pm 0.1$. The solution quality of these small structures is sensitive to the grid resolution, thus the AMR technique can play an important role to build the accurate and efficient numerical models for MHD flows.

\subsubsection{Blast wave in a magnetic field}

A two-dimensional blast wave test was carried out in \cite{Skinner_2010,Jiang_2010} to check the models' performance when dealing with MHD shock waves affected by a strong magnetic field. The initial conditions are given as a static plasma with the uniform density $\rho=1$ and a magnetic field $\left(B_x,B_y,B_z\right)=\left(\frac{\sqrt{2}}{2},\frac{\sqrt{2}}{2},0\right)$ within a domain $\left(x,y\right)\in\left[1,2\right]\times\left[-0.5,0.5\right]$. The pressure distribution is specified to trigger the blast wave as
\begin{equation}
p=\left\{\begin{array}{ll}
10&{\rm if} \sqrt{\left(x-1.5\right)^2+y^2} < 0.1,\\
0.1&{\rm otherswise}.
\end{array}\right.
\end{equation}
and the ratio of specific heats $\gamma=\frac{5}{3}$ is used in this test.

\begin{figure}[htbp]
\centering
\subfigure[Density]
{\includegraphics[width=0.45\textwidth]{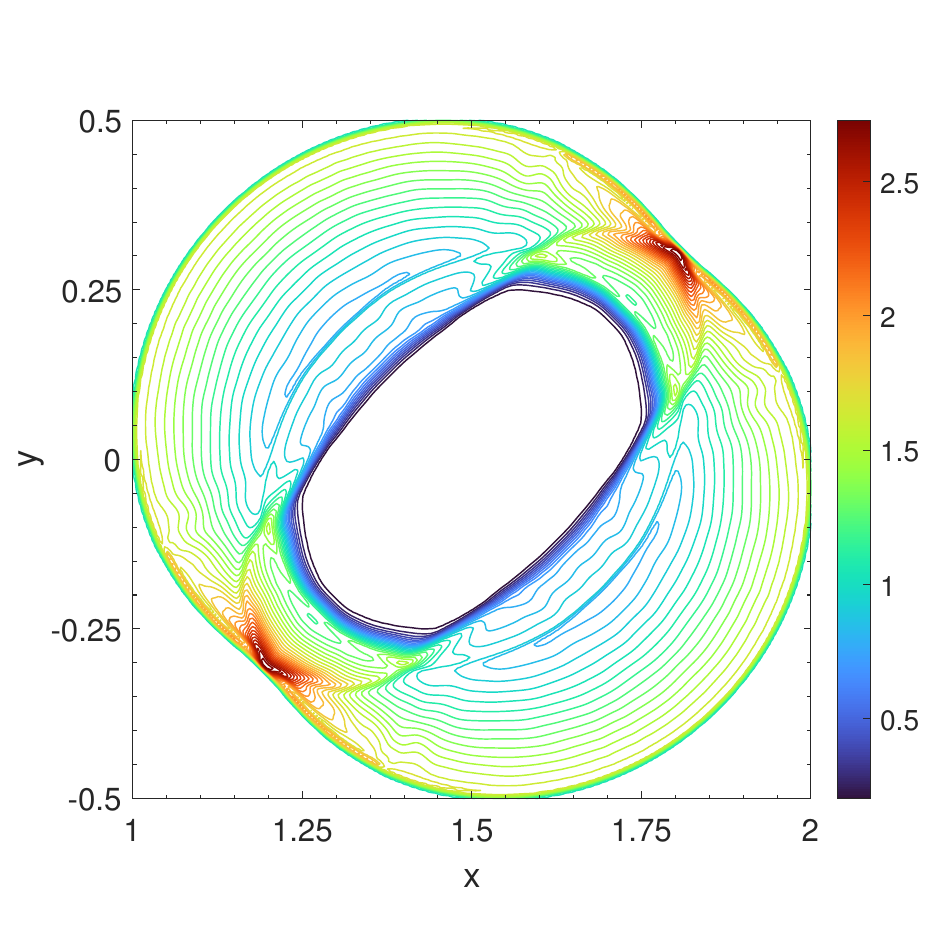}}
\subfigure[Pressure]
{\includegraphics[width=0.45\textwidth]{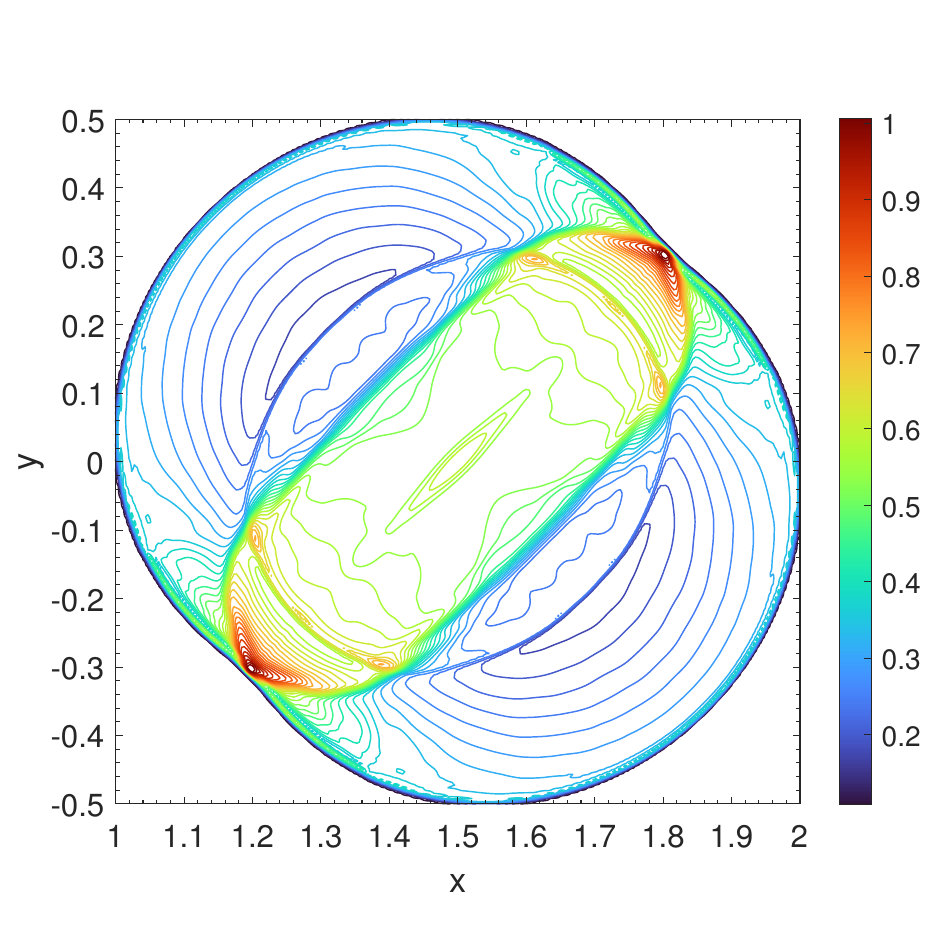}}\\
\subfigure[Magnetic pressure]
{\includegraphics[width=0.45\textwidth]{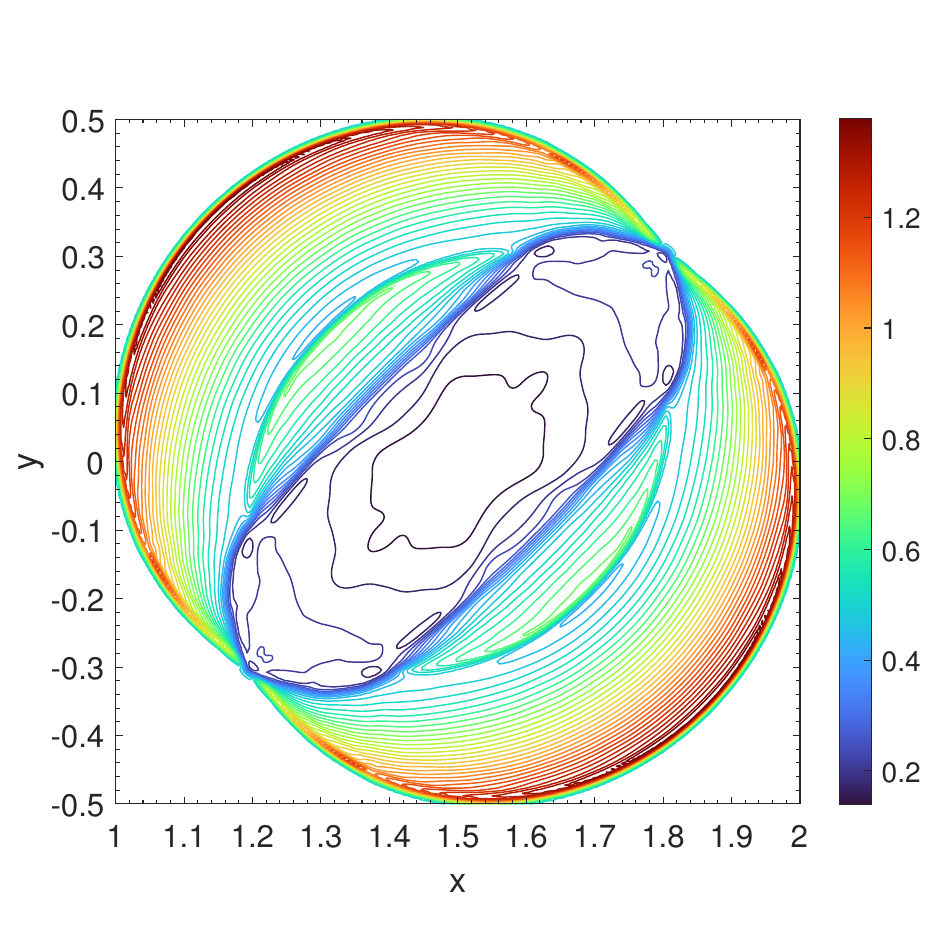}}
\subfigure[Specific kinetic energy]
{\includegraphics[width=0.45\textwidth]{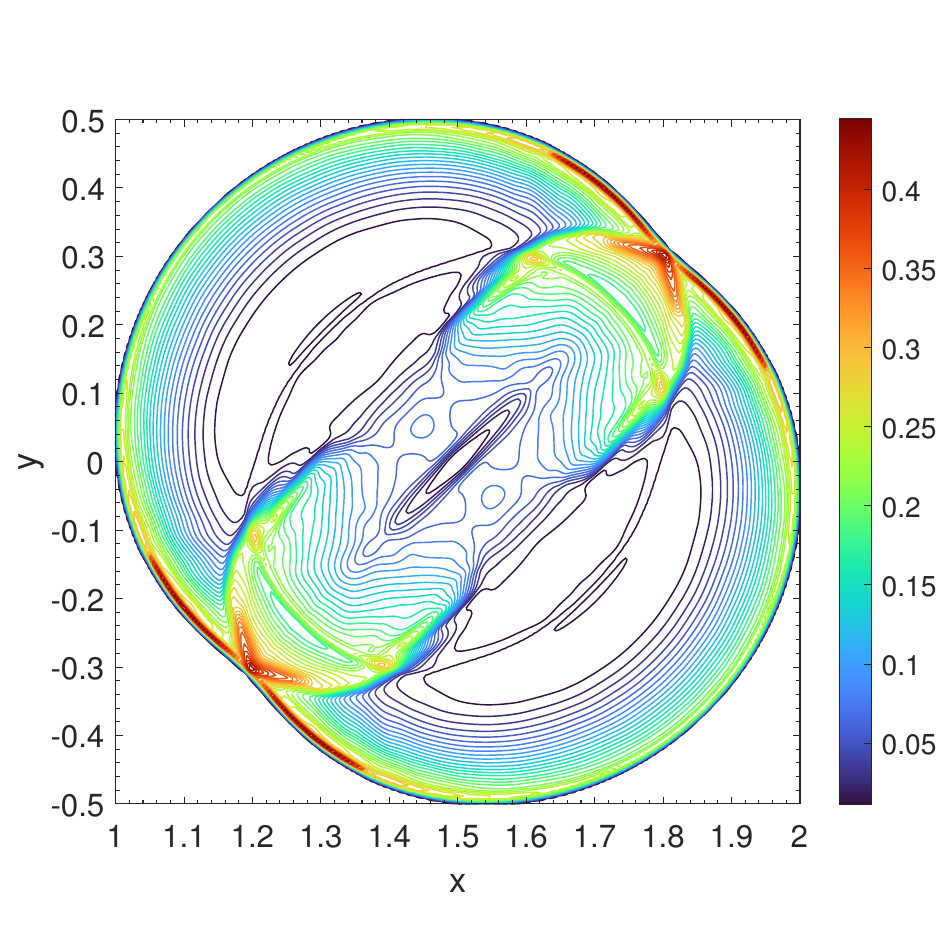}}\\
\caption{Contour plots of numerical results of blast wave problem at $t=0.2$ (N=256) by two-step BVD scheme.}\label{blastwaveresults}
\end{figure}

\begin{figure}[htbp]
\centering
\centering
\subfigure[Density]
{\includegraphics[width=0.45\textwidth]{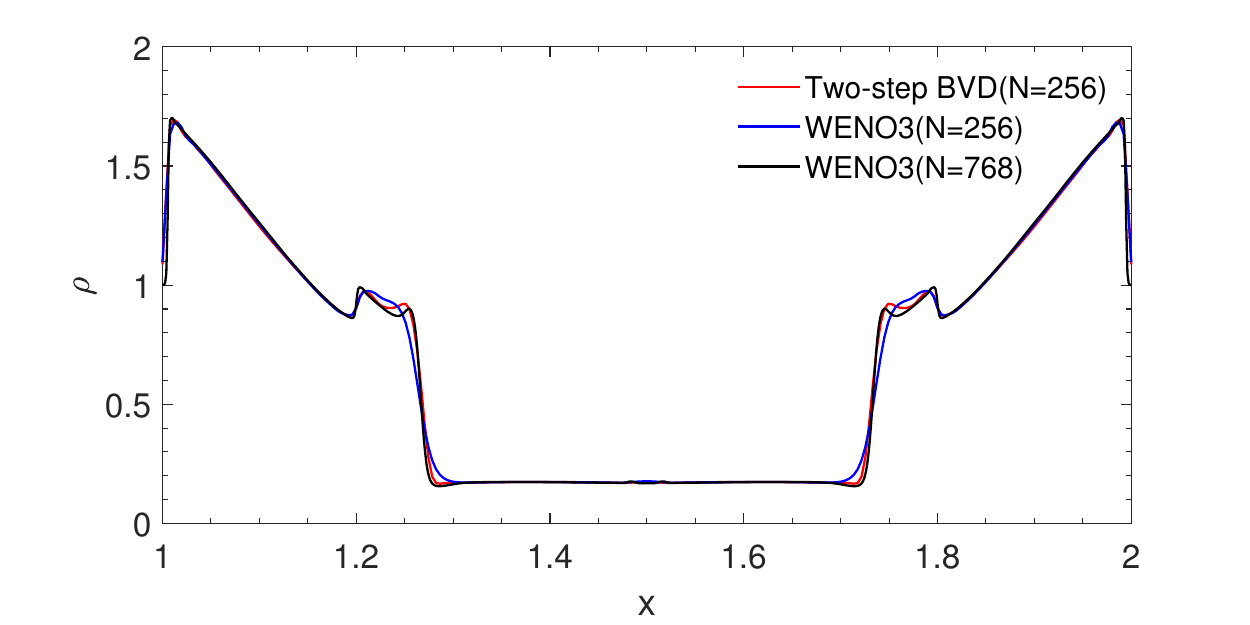}}
\subfigure[Pressure]
{\includegraphics[width=0.45\textwidth]{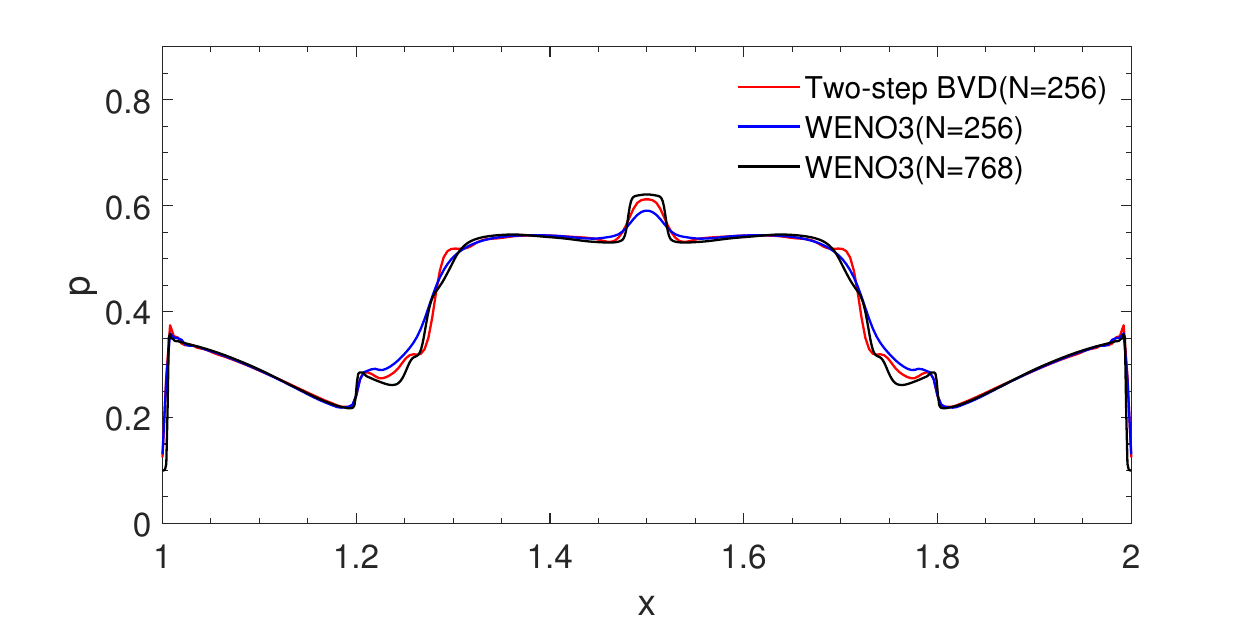}}\\
\subfigure[Magnetic pressure]
{\includegraphics[width=0.45\textwidth]{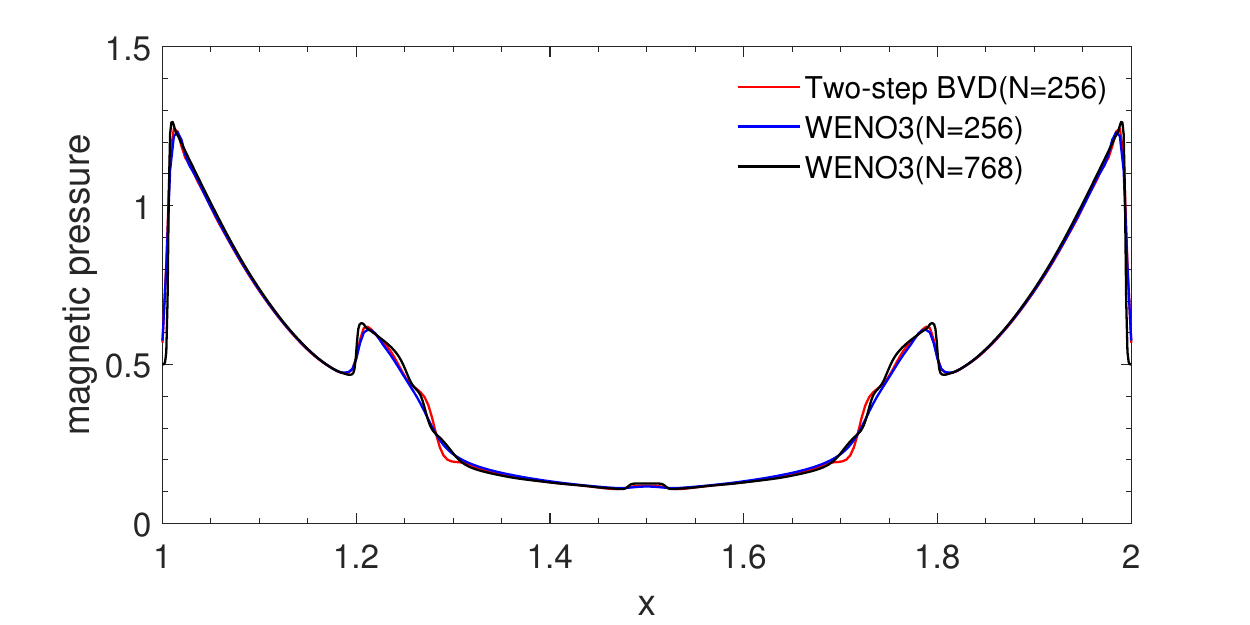}}
\subfigure[Specific kinetic energy]
{\includegraphics[width=0.45\textwidth]{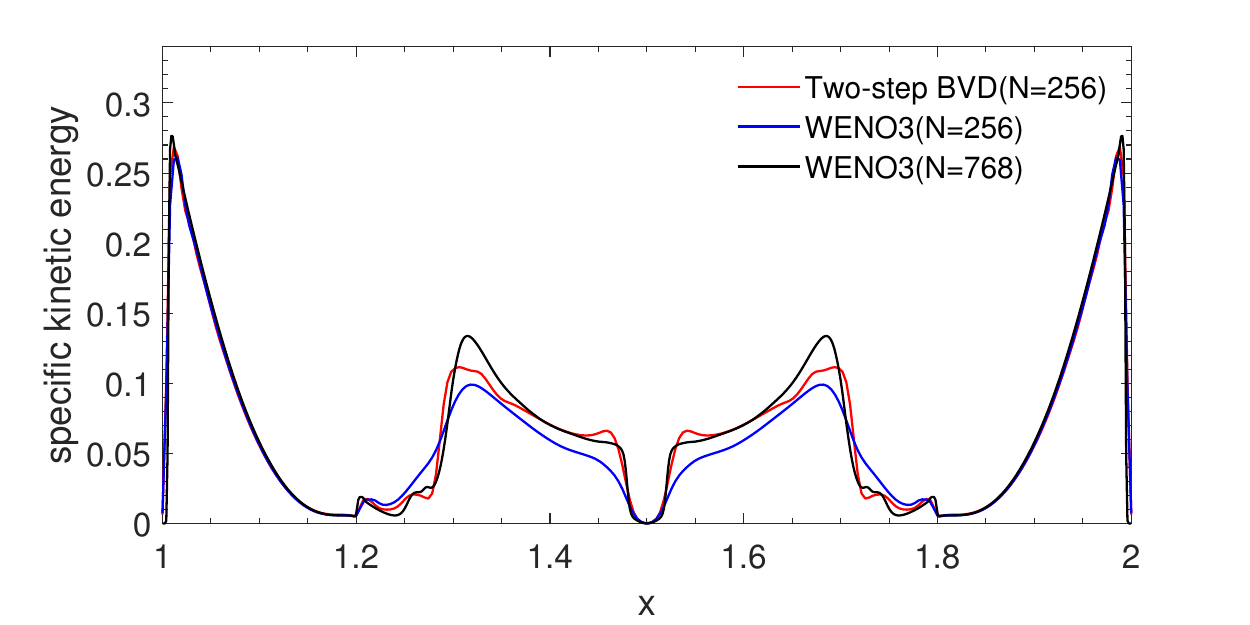}}\\
\caption{Cross-section plots of numerical results of blast wave problem along line $y=0$.}\label{blastwaveresults2}
\end{figure}

We give the numerical results by two-step BVD scheme in Fig. \ref{blastwaveresults} for contour plots of the density, pressure, magnetic pressure and specific kinetic energy. We use $256\times256$ cells and the numerical model is integrated for 0.2s. In comparison with the reference solution in  \cite{Skinner_2010,Jiang_2010,Yang_2018}, our BVD scheme accurately capture the MHD shocks without spurious oscillations and the multi-scale flow structures agree well. As in the previous tests, we also quantitatively evaluate the performance of the proposed scheme by comparing the results by BVD, WENO3 and WENO3 with triple resolution (N=768) in a cross-section plot along line $y=0$ (Fig. \ref{blastwaveresults2}). On the same grid, BVD scheme is more accurate than WENO3 scheme in reproducing some small-scale distributions. The proposed BVD scheme shows better solution fidelity in this blast wave test once again.

\section{Conclusion}\label{sec:conclusions}

MHD flows consist of the physical phenomena over a wide range of temporal and spatial scales. Thus, it is a very challenging work to develop a high-fidelity MHD model, which can capture the shock waves without non-physical oscillations while preserving the high-order accuracy to reproduce the complex vortex structures. In this study, we develop a new finite volume scheme for MHD equations based on the BVD algorithm. Two kinds of alternative functions, including a quadratic polynomial and a hyperbolic tangent function, are considered for the spatial reconstructions and the optimal one is chosen by minimizing the total boundary variation of each computational cell. The accuracy of proposed numerical scheme in simulating MHD flows has been verified by the widely used benchmark tests including both smooth and non-smooth solutions in comparison with a linear upwind scheme and a WENO scheme using the same stencil. To implement an accurate and efficient numerical model for practical applications involving MHD flows, we tend to focus on the development of approximate Riemann solvers suited for MHD equations, high-order non-divergent correction algorithms and the application of adaptive mesh refinement technique in the future.

%\section*{Acknowledgments}

%This work is supported by

\bibliography{refs}

%\printbibliography{}

\clearpage

\listoftables

\clearpage

\begin{table}[htbp]
\caption{Errors and convergence rates of advection of a sine wave at $t=2$.}\label{accuracy}
\centering
\begin{tabular}{cccccc} \hline
Schemes & mesh & $l_{1}$ errors & $l_{1}$ orders & $l_{\infty}$ errors & $l_{\infty}$ orders \\
\hline
         & 20  & 1.003E-01  &   	   & 1.423E-01  &       \\
         & 40  & 3.122E-02  & 1.68  & 5.671E-02  & 1.33  \\
WENO3    & 80  & 7.055E-03  & 2.15  & 2.031E-02  & 1.48  \\
         & 160 & 7.310E-04  & 3.27  & 3.984E-03  & 2.35  \\
         & 320 & 4.234E-05  & 4.11  & 3.830E-04  & 3.38  \\ \hline
         & 20  & 1.635E-02  &       & 1.646E-02  &       \\
One-step & 40  & 2.084E-03  & 2.97  & 2.087E-03  & 2.98  \\
BVD      & 80  & 2.615E-04  & 2.99  & 2.616E-04  & 3.00  \\
         & 160 & 3.271E-05  & 3.00  & 3.272E-05  & 3.00  \\
         & 320 & 4.090E-06  & 3.00  & 4.090E-06  & 3.00  \\ \hline
         & 20  & 1.635E-02  &       & 1.646E-02  &       \\
Two-step & 40  & 2.084E-03  & 2.97  & 2.087E-03  & 2.98  \\
BVD      & 80  & 2.615E-04  & 2.99  & 2.616E-04  & 3.00  \\
         & 160 & 3.271E-05  & 3.00  & 3.272E-05  & 3.00  \\
         & 320 & 4.090E-06  & 3.00  & 4.090E-06  & 3.00  \\ \hline
         & 20  & 1.635E-02  &       & 1.646E-02  &       \\
         & 40  & 2.084E-03  & 2.97  & 2.087E-03  & 2.98  \\
Upwind3  & 80  & 2.615E-04  & 2.99  & 2.616E-04  & 3.00  \\
         & 160 & 3.271E-05  & 3.00  & 3.272E-05  & 3.00  \\
         & 320 & 4.090E-06  & 3.00  & 4.090E-06  & 3.00  \\ \hline			
\end{tabular}
\end{table}

\clearpage

\begin{table}[htbp]
\caption{ Errors and convergence rates of 2D circularly polarized Alfv\'{e}n wave at time $t= 2$.}\label{2D_MHD_accuracy}
\centering
\begin{tabular}{ccccccccccc}\hline
\multirow{2}{*}{Schemes} & \multirow{2}{*}{mesh} & \multicolumn{2}{c}{$u$} & \multicolumn{2}{c}{$v$} & \multicolumn{2}{c}{$B_x$} & \multicolumn{2}{c}{$B_y$} \\
\cmidrule(r){3-4} \cmidrule(r){5-6} \cmidrule(r){7-8} \cmidrule(r){9-10}
& & $l_{1}$ error & order & $l_1$ error & order & $l_1$ error & order & $l_1$ error & order \\ \hline
& 10 & 3.128E-02 & & 6.256E-02 & & 9.409E-02 & & 1.256E-01 & \\
& 20 & 8.289E-03 & 1.92 & 1.658E-02 & 1.92 & 2.504E-02 & 1.91 & 3.349E-02 & 1.91	\\
WENO3& 40 & 3.251E-03 & 1.35 & 6.503E-03 & 1.35 & 9.761E-03 & 1.36 & 1.302E-02 & 1.36	\\
& 80 & 5.876E-04 & 2.47 & 1.175E-03 & 2.47 & 1.746E-03 & 2.48 & 2.317E-03 & 2.49	\\ \hline

& 10 & 1.149E-02 & & 2.298E-02 & & 3.452E-02 & & 4.605E-02 & \\
One-step& 20 & 1.667E-03 & 2.79 & 3.333E-03 & 2.79 & 5.002E-03 & 2.79 & 6.671E-03 & 2.79	\\
BVD& 40 & 2.110E-04 & 2.98 & 4.221E-04 & 2.98 & 6.333E-04 & 2.98 & 8.445E-04 & 2.98	\\
& 80 & 2.631E-05 & 3.00 & 5.263E-05 & 3.00 & 7.896E-05 & 3.00 & 1.053E-04 & 3.00	\\ \hline

& 10 & 6.935E-03 & & 1.384E-02 & & 2.081E-02 & & 2.779E-02 & \\
Two-step& 20 & 1.667E-03 & 2.06 & 3.334E-03 & 2.05 & 5.002E-03 & 2.06 & 6.671E-03	& 2.06	\\
BVD& 40 & 2.110E-04 & 2.98 & 4.221E-04 & 2.98 & 6.334E-04 & 2.98 & 8.445E-04	& 2.98	\\
& 80 & 2.631E-05 & 3.00 & 5.263E-05 & 3.00 & 7.896E-05 & 3.00 & 1.053E-04	& 3.00	\\ \hline

& 10 & 1.149E-02 & & 2.298E-02 & & 3.452E-02 & & 4.605E-02 & \\
& 20 & 1.667E-03 & 2.79 & 3.333E-03 & 2.79 & 5.002E-03 & 2.79 & 6.671E-03	& 2.79	\\
Upwind3& 40 & 2.110E-04 & 2.98 & 4.221E-04 & 2.98 & 6.333E-04 & 2.98 & 8.445E-04	& 2.98	\\
& 80 & 2.631E-05 & 3.00 & 5.263E-05 & 3.00 & 7.896E-05 & 3.00 & 1.053E-04	& 3.00	\\ \hline

\end{tabular}
\end{table}

\clearpage

\listoffigures

\clearpage

\clearpage

\clearpage

\clearpage

\clearpage

\clearpage

\clearpage

\clearpage

\clearpage

\clearpage

\end{document}